\documentclass[12pt]{article}
\usepackage{lipsum}

\usepackage{xr-hyper}
\usepackage{amsthm}
\usepackage{fullpage}
\usepackage{footnote}
\usepackage{tabu}
\usepackage{booktabs}
\usepackage[title]{appendix}
\usepackage{scribe}
\usepackage{caption}
\usepackage{setspace}
\usepackage{xcolor}

\definecolor{comcolor}{rgb}{0,0.5,0}
\definecolor{funcolor}{rgb}{0,0.4,1}
\definecolor{concolor}{rgb}{0,0,1}

\definecolor{codegreen}{rgb}{0,0.6,0}
\definecolor{codegray}{rgb}{0.5,0.5,0.5}
\definecolor{codepurple}{rgb}{0.58,0,0.82}
\definecolor{backcolour}{rgb}{0.93,0.93,0.93}
\newcommand{\funcol}[1]{\textcolor{funcolor}{#1}}

\newcommand{\comcolor}[1]{\textcolor{comcolor}{#1}}
\newcommand{\opcol}[1]{\textcolor{red}{#1}}

\newcommand{\strcol}[1]{\textcolor{codepurple}{#1}}

\providecommand{\keywords}[1]
{
  \small	
  \textbf{\textit{Keywords---}} #1
}
\usepackage{etoolbox}

\makeatletter
\patchcmd{\@makecaption}
  {\parbox}
  {\advance\@tempdima-\fontdimen2} 
  {}{}
\makeatother 

\usepackage{titling} 

\usepackage{amsmath,bbm,amssymb, amsfonts}
\usepackage{mathtools}
\usepackage{graphicx,psfrag,epsf}
\usepackage{enumitem}
\usepackage{natbib}
\usepackage{url} 
\usepackage{color}
\usepackage{algorithmicx}
\usepackage{booktabs}
\usepackage{siunitx}
\usepackage{array}
\usepackage{rotating}

\usepackage{adjustbox}

\usepackage{algorithm}
\usepackage{algpseudocode}

\usepackage{tikz}

\usepackage{url}

\usepackage{multicol}
\usepackage{xcolor}
\usepackage{algorithm}

\makeatletter
\def\breve{\mathpalette\wide@breve}
\def\wide@breve#1#2{\sbox\z@{$#1#2$}%
	\mathop{\vbox{\m@th\ialign{##\crcr
				\kern0.08em\brevefill#1{0.8\wd\z@}\crcr\noalign{\nointerlineskip}%
				$\hss#1#2\hss$\crcr}}}\limits}
\def\brevefill#1#2{$\m@th\sbox\tw@{$#1($}%
	\hss\resizebox{#2}{\wd\tw@}{\rotatebox[origin=c]{90}{\upshape(}}\hss$}
\makeatletter

\def\gobblestop#1#2{#1}
\def\killstop{%
	\aftergroup\gobblestop
}

\def\thick#1{\hbox{\rlap{$#1$}\kern0.25pt\rlap{$#1$}\kern0.25pt$#1$}}

\def\smbalpha{\boldsymbol{{\scriptstyle{\alpha}}}}

\def\smbalpha{\widehat{\smbalpha}}

\def\hbar{\bar{ h}}

\def\mybox#1{\vskip1mm \begin{center}
        \hspace{.0\textwidth}\vbox{\hrule\hbox{\vrule\kern6pt
\parbox{.9\textwidth}{\kern6pt#1\vskip6pt}\kern6pt\vrule}\hrule}
        \end{center} \vskip-5mm}
\def\lboxit#1{\vbox{\hrule\hbox{\vrule\kern6pt
      \vbox{\kern6pt#1\vskip6pt}\kern6pt\vrule}\hrule}}

\def\thickboxit#1{\vbox{{\hrule height 1mm}\hbox{{\vrule width 1mm}\kern6pt
          \vbox{\kern6pt#1\kern6pt}\kern6pt{\vrule width 1mm}}
               {\hrule height 1mm}}}

\def\fat#1{\hbox{\rlap{$#1$}\kern0.25pt\rlap{$#1$}\kern0.25pt$#1$}}

\newcolumntype{R}{@{\extracolsep{0.5cm}}r@{\extracolsep{0pt}}}%
\newcolumntype{E}{@{\extracolsep{0.25cm}}c@{\extracolsep{0pt}}}%

\makeatletter
\newcommand{\distas}[1]{\mathbin{\overset{#1}{\kern\z@\sim}}}%
\makeatother

\usepackage{xr-hyper}
\usepackage[colorlinks, linkcolor=blue, anchorcolor=blue, citecolor=blue]{hyperref}

\usepackage[left=1in,right=1in,top=1in,bottom=1in]{geometry}

\makeatletter
\newcommand*{\addFileDependency}[1]{
  \typeout{(#1)}
  \@addtofilelist{#1}
  \IfFileExists{#1}{}{\typeout{No file #1.}}
}
\makeatother

\newcommand{\blind}{1}

\begin{document}

\if1\blind
{
	\title{\bf Tutorial on Bayesian Functional Regression Using Stan}
	\author{Ziren Jiang$^{1}$, Ciprian Crainiceanu$^{2}$, Erjia Cui$^{1}$\thanks{corresponding author: ecui@umn.edu}\\ [6pt]
		$^{1}$Department of Biostatistics and Health Data Science, University of Minnesota \\
        $^{2}$Division of Biostatistics, Johns Hopkins University \\ [8pt]
	}
	\date{}
	\maketitle
} \fi

\if0\blind
{
	\title{\bf Estimating causal effects of functional treatments with modified functional treatment policies}
	\date{}
	\maketitle
} \fi


\begin{abstract}
This manuscript provides step-by-step instructions for implementing Bayesian functional regression models using Stan. Extensive simulations indicate that the inferential performance of the methods is comparable to that of state-of-the-art frequentist approaches. However, Bayesian approaches allow for more flexible modeling and provide an alternative when frequentist methods are not available or may require additional development. Methods and software are illustrated using the accelerometry data from the National Health and Nutrition Examination Survey (NHANES).

\keywords{
Bayesian data analysis; \;Functional data analysis; \; Stan; \; Functional principal component analysis; \; Functional Cox regression;
}
\end{abstract}%

\newpage
\singlespacing
\setlength{\abovedisplayskip}{7pt}%
\setlength{\belowdisplayskip}{7pt}%
\setlength{\abovedisplayshortskip}{5pt}%
\setlength{\belowdisplayshortskip}{5pt}%
\section{Introduction}\label{sec:1}
The collection of ever more complex and high-dimensional data in many scientific areas has led to increased interest in Functional Data Analysis (FDA) \color{black} - a branch of statistics that focuses on the analysis of data that can be represented as functions, curves, or trajectories observed over a continuum, such as time or space \color{black} \citep{ramsaysilv2005, kokoszka2017introduction, crainiceanu2024functional}. Some examples include continuous glucose monitoring \citep{sergazinov2023case}, colon carcinogenesis \citep{baladandayuthapani2008bayesian}, electroencephalography \citep{di2009multilevel}, epidemiological monitoring \citep{salvatore2015wastewater}, fiber photometry \citep{loewinger2024statistical}, genomics \citep{leng2006classification}, house pricing \citep{peng2014time}, neuroimaging \citep{goldsmith2011penalized, goldsmith2012longitudinal, reiss2010functional, cui2022fast, li2022fixed}, online auction \citep{liu2009estimating}, polysomnography \citep{ballard2024functional}, traffic flow \citep{chiou2012dynamical}, and wearable and implantable devices \citep{khan2016monitoring, cui2023fast}, among others.

The complexity, size, and structure of the new data sets require the development of new functional analytic methods and, especially, software that can be used, adapted, and tested in reasonable time. In a recent monograph, Crainiceanu et al. \cite{crainiceanu2024functional} showed that functional regression models can be viewed as mixed effects models and existing frequentist software, such as the \texttt{refund} \citep{goldsmith_refund_2024} and \texttt{mgcv} \citep{wood2001mgcv} packages in \texttt{R} \citep{R}, can be used to fit functional regression models.

\color{black}Bayesian analysis naturally accommodates hierarchical structures, allows for the incorporation of prior information, and provides a coherent framework for quantifying uncertainty in both fixed and random effects, making it especially well suited for mixed effect models. \color{black}
This suggests that Bayesian analyses may provide a powerful alternative to frequentist approaches that could be adapted faster to emerging data structures. \color{black}In this paper, we provide a tutorial on scalar-on-function regression (SoFR, where the outcome is scalar and predictors include functional variables), function-on-scalar regression (FoSR, where the outcome is a functional variable and predictors are scalar), and functional Cox regression (for time-to-event outcomes with functional predictors), all using Bayesian methods implemented in \texttt{Stan}. \color{black} \citep{carpenter2017stan}. This type of models has been introduced by multiple researchers using different nomenclatures  \cite{ramsaysilv2005,Tikhonov63,wahba1990}; the first use of the SoFR/FoSR nomenclature can be traced to Reiss et al.\cite{reissfosr}. The first Bayesian implementation of these models was provided by  Crainiceanu and Goldsmith
\cite{crainiceanu2010bayesian} using \texttt{WinBUGS} \citep{lunn2000winbugs}. The current tutorial will substantially expand existing Bayesian functional regression software by taking advantage of the rapid development of Bayesian computation in the past $15$ years. In particular, we: (1) provide a detailed description of Bayesian functional regression models and the corresponding \texttt{Stan} implementation with step-by-step code demonstrations; 
(2) propose a Bayesian joint model to incorporate functional principal component analysis (FPCA) into functional regression models and estimate the functional coefficients and principal component scores simultaneously; (3) conduct extensive simulation experiments to validate the software and compare it with the state-of-the-art frequentist software; and (4) provide an \texttt{R} package, \texttt{refundBayes}, for model implementation. \color{black} The supplementary R Markdown file contains the complete code based on our \texttt{refundBayes} package. \color{black} Some familiarity with splines, regression, and penalized approaches is assumed; see Crainiceanu et al.\cite{crainiceanu2024functional} for an introduction, especially Chapters 4-6. 


The rest of the manuscript is organized following the logic of functional regression. In Section \ref{sec:2}, we introduce the Bayesian scalar-on-function regression model and its implementation in \texttt{Stan}. In Section \ref{sec:3}, we introduce the Bayesian functional Cox regression model for time-to-event outcome. Both Sections \ref{sec:2} and \ref{sec:3} use the observed functional covariate as the predictor variable. In Section \ref{sec:4}, we illustrate how to incorporate FPCA into our Bayesian functional regression model. In Section \ref{sec:5}, we introduce the Bayesian function-on-scalar regression model. Section \ref{sec:6} provides real-world application examples using the NHANES data. We conclude with discussions in Section \ref{sec:7}.

\section{Bayesian Scalar-on-Function Regression}\label{sec:2}

Penalized splines have become the practical standard in semiparametric regression \citep{marxeilers1996,osullivan1986,ruppert2003semiparametric,wood2001mgcv,wood2016smoothing} because they provide an excellent balance between computational complexity, adaptation to real data scenarios, and inferential capabilities. Penalized splines  use a moderately large number of basis functions to account for the maximum complexity of the model and quadratic penalties to control the smoothness of the fit. By showing that these penalized models are equivalent to mixed effects models, software that was originally developed for inference in mixed effects models can be expanded to semiparametric regression. These ideas have been extended to functional regression  that also have a mixed effects representation \cite{crainiceanu2010bayesian,goldsmith2011penalized}. In this section, we present the \texttt{Stan} program for implementing a Bayesian Scalar-on-Function Regression (SoFR) model, where functional parameters are modeled nonparametrically using penalized splines. 
The inferential performance of these models compared with existing frequentist models are assessed via simulations.

\subsection{The SoFR Model}\label{sec:2.1}
We start by introducing the data structure for SoFR. For subject $i=1,\ldots,n$, let $Y_i$ be the outcome, $\mathbf{Z}_i$ be the $p\times 1$ dimensional vector of scalar predictors, and $\{W_i(t_{im}), t_{im}\in [0,1]\}$ with $m=1,\ldots,M_i$ be a functional predictor, where $M_i$ is the number of functional observations for study participant $i$. Here, we assume that the observation time points are identical for all subjects, such that $M_1=M_2=...=M_n=M$ and $t_{im}=t_m$ for all $i=1,...,n$ and $m=1,...,M_i$. For data observed at irregular locations, users can refer to Section $\ref{sec:4}$, where we introduce the Bayesian joint modeling of functional principal component analysis (FPCA) and functional regression. Although we use a single functional predictor here to illustrate the model, the proposed method and software can be easily extended to multiple functional predictors.

The SoFR model assumes that the distribution of $Y_i$ follows an exponential family with mean $\mu_i$, and the linear predictor $\eta_i = g(\mu_i)$ has the following structure 
\begin{equation}\label{eq:linearpred}
    \eta_i=\eta_0+\int_0^1 W_i(t)\beta(t)dt + \mathbf{Z}^t_i\boldsymbol{\gamma}\;, 
\end{equation}
where $\eta_0$ is the overall intercept, $\beta(\cdot)\in L_2[0,1]$ is the functional coefficient, and $\boldsymbol{\gamma}$ is a $p\times 1$ dimensional vector of parameters. If $\psi_1(t),...,\psi_K(t)$ is a collection of $K$ pre-specified basis functions and $\beta(t)=\sum_{k=1}^Kb_k\psi_k(t)$, the linear predictor can be re-written as:
\begin{equation}\label{eq:linearpredmat}
    \begin{split}
        \eta_i&=\eta_0+\sum_{k=1}^Kb_k\int_0^1 W_i(t)\psi_k(t)dt + \mathbf{Z}^t_i\boldsymbol{\gamma}\\
       &\approx \eta_0+\sum_{k=1}^Kb_k  \sum_{m=1}^M L_{m}W_i(t_m)\psi_k(t_m)+\mathbf{Z}^t_i\boldsymbol{\gamma}\\
        &=\eta_0+ \mathbf{X}_i^t\boldsymbol{b}+\mathbf{Z}_i^t\boldsymbol{\gamma}\;,
    \end{split}
\end{equation}
where $L_{m}=t_{m+1}-t_m$, and the approximation sign on the second line indicates the Riemann sum approximation to the integral. The $K\times 1$ dimensional vector $\mathbf{X}_i=(X_{i1},...,X_{iK})^t$ has the $k$-th entry equal to
\begin{equation}\label{eq:xmat}
    X_{ik}=\sum_{m=1}^M L_{m}W_i(t_m)\psi_k(t_m)\;.
\end{equation}
 Let $\mathbf{X}=[\mathbf{X}_1,...,\mathbf{X}_n]$ be the $K\times n$ matrix of functional covariates, and $\mathbf{Z}=[\mathbf{Z}_1,...,\mathbf{Z}_n]$ be the $p\times n$ matrix of scalar covariates. The linear predictor $\boldsymbol{\eta}=(\eta_1,...,\eta_n)^t$ can then be expressed as
\begin{equation}\label{eq:theta}
    \boldsymbol{\eta}= \eta_0\boldsymbol{J}_n+\mathbf{X}^t\boldsymbol{b}+\mathbf{Z}^t\boldsymbol{\gamma}\;,
\end{equation}
where $\boldsymbol{J}_n$ is the $n\times1$ vector with all entries equal to $1$. Note that this, unpenalized, model is a standard generalized linear model (GLM). Smoothness can be induced on the functional regression coefficient, $\beta(t)$, by penalizing the spline coefficients, $\boldsymbol{b}$. In the next section we show that this is equivalent to assuming a particular prior on $\boldsymbol{b}$, which transforms the model into a specific generalized linear mixed effects model (GLMM).

\subsection{Incorporating Penalized Splines}\label{sec:2.2}
A common approach to induce smoothness on $\beta(t)$ is to assume that the number of spline basis functions, $K$, is relatively large and add a quadratic penalty on the regression coefficients. We focus on penalties on the integral of the square of the second derivative, $\int \{\beta''(t)\}^2dt$. This penalty was introduced by Grace Wahba and collaborators  \citep{cravenwahba1979,kimeldorfwahba1970,Wahba1983}  for smoothing spline regression (as many basis functions as observations) and by Finbarr O'Sullivan for penalized spline regression (B-splines with a smaller number of knots than the number of observations)  \citep{osullivan1986}. Given its historical and practical importance, we would like to coin this foundational concept as the Wahba-O'Sullivan smoothing penalty. While we focus on this penalty here, the methods described can be applied to any other type of quadratic penalty and/or basis functions.

The Wahba-O'Sullivan penalty can be re-written as
\begin{equation}
    \begin{split}
        \int \{\beta''(t)\}^2dt&=\int \{\sum_{k=1}^Kb_k\psi_k''(t)\}^2dt \\
        &= \int\{\{\boldsymbol{\psi''\textnormal{(t)}}\}^t\boldsymbol{b}\}^t\{\{\boldsymbol{\psi''\textnormal{(t)}}\}^t\boldsymbol{b}\}dt\\
        &=\boldsymbol{b}^t \int \{\boldsymbol{\psi}''\textnormal{(t)}\}\{\boldsymbol{\psi}''\textnormal{(t)}\}^t dt \boldsymbol{b}=\boldsymbol{b}^t\mathbf{S}\boldsymbol{b}\;,
    \end{split}
\end{equation}
where $\mathbf{S}=\int \{\boldsymbol{\psi}''\textnormal{(t)}\}\{\boldsymbol{\psi}''\textnormal{(t)}\}^t dt$ is the penalty matrix and $\boldsymbol{b}=(b_1,...,b_K)^t$ are spline coefficients. A penalized spline regression approach would minimize the criterion 
\begin{equation}-2\log L(\mathbf{Y}|\mathbf{X},\mathbf{Z},\eta_0,\boldsymbol{b},\boldsymbol{\gamma})+\lambda\boldsymbol{b}^t\mathbf{S}\boldsymbol{b}\;,
\label{eq:penalized_like}
\end{equation}
where $\lambda$ is a scalar non-negative smoothing parameter that controls the complexity of the spline fit. In a Bayesian modeling context, the spline coefficients, $\boldsymbol{b}$, can be viewed as random variables with a multivariate normal prior  
\begin{equation}\label{eq:penaltyb}
    p(\boldsymbol{b})\propto \exp\left(-\frac{\boldsymbol{b}^t\mathbf{S}\boldsymbol{b}}{\sigma^2_b}\right)\;,
\end{equation}
where $\sigma^2_b$ is the parameter that controls the smoothness of the functional coefficient $\beta(t)$.

The prior distribution in~\eqref{eq:penaltyb} is easy to write, but its implementation may be subject to numeric instability, especially when multiple functional and scalar predictors are considered. To address this problem we use a linear reparametrization of the model parameters, $\boldsymbol{b}$, that corresponds to independent normal priors on the transformed parameters. This method was  introduced by  Ruppert et al. \cite{ruppert2003semiparametric} in the context of semiparametric regression and was first implemented in a Bayesian context by  Crainiceanu et al. \cite{crainiceanuwinbugs} for penalized thin plate splines. The method was likely developed in parallel and extended by Wood and collaborators \cite{wood2006,wood2001mgcv,wood2015package} in the context of multiple functions and penalty matrices that are not of full rank. Here we use a description of the method provided by Scheipl et al. \cite{scheipl2012spike} in the context of Bayesian analysis of additive regression models; the method was also used in the context of function-on-scalar Bayesian inference by Sun and Kowal \cite{Sun22072024}.  

Consider the spectral decomposition $\mathbf{S}=\mathbf{U}\mathbf{VU}^t$, where  $\mathbf{U}$ is a $K\times K$ dimensional matrix such that $\mathbf{U}^t\mathbf{U}=\mathbf{U}\mathbf{U}^t=\mathbf{I}_K$ and $\mathbf{V}$ is a diagonal $K\times K$ dimensional matrix. The columns of the matrix $\mathbf{U}$ are the eigenvectors of the penalty matrix $\mathbf{S}$, while the diagonal elements of the matrix $\mathbf{V}$ are its corresponding eigenvalues. If $K_0$ is the rank of $\mathbf{S}$, we re-write $\mathbf{S}=[\mathbf{U}_+ \mathbf{U}_0]\big[\begin{smallmatrix}
  \mathbf{V}_+ & \mathbf{0}\\
  \mathbf{0} & \mathbf{0}
\end{smallmatrix}\big][\mathbf{U}_+ \mathbf{U}_0]^t$, where $\mathbf{U}_+$ is the $K\times K_0$ matrix of eigenvectors associated with strictly positive eigenvalues, $\mathbf{U}_0$ is the $K\times (K-K_0)$ matrix of eigenvectors associated with the zero eigenvalues, and $\mathbf{V}_+$ is the $K_0\times K_0$ dimensional diagonal matrix of non-zero eigenvalues. Consider the following reparametrization of the spline coefficients
\begin{equation}\label{eq:transb}
\widetilde{\boldsymbol{b}}=\widetilde{\mathbf{V}}^{1/2}\mathbf{U}^t\boldsymbol{b}\;,
\end{equation}
where $\widetilde{\mathbf{V}}^{1/2}=\big[\begin{smallmatrix}
  \mathbf{V}_+^{1/2} & \mathbf{0}\\
  \mathbf{0} & \mathbf{I}_{K-K_0}
\end{smallmatrix}\big]$,  $\mathbf{V}_+^{1/2}$ is the diagonal matrix of square roots of the non-zero eigenvalues, and  $\mathbf{I}_{K-K_0}$ is the identity matrix of rank $K-K_0$. With this notation we also have $\boldsymbol{b}=\mathbf{U}\widetilde{\mathbf{V}}^{-1/2}\widetilde{\boldsymbol{b}}$, which provides a way of recovering the original spline parameters, $\boldsymbol{b}$, from the transformed parameters, $\widetilde{\boldsymbol{b}}$. This is important because working with $\widetilde{\boldsymbol{b}}$ is easier, but reconstructing the functional parameter $\beta(t)$ requires $\boldsymbol{b}$.

We further partition $\widetilde{\boldsymbol{b}}$ into $\widetilde{\boldsymbol{b}}=(\widetilde{\boldsymbol{b}}_r^t,\widetilde{\boldsymbol{b}}_f^t)^t$ where $\widetilde{\boldsymbol{b}}_r$ are the first $K_0$ coefficients. From the definition, $\widetilde{\boldsymbol{b}}_r = \mathbf{[V^{1/2}_{+}\  0]}\mathbf{U}^t\boldsymbol{b}$ and
\begin{equation}
\widetilde{\boldsymbol{b}}_r^t\mathbf{I}_{K_0}\widetilde{\boldsymbol{b}}_r = \boldsymbol{b}^t\mathbf{U ([V^{1/2}_{+}\  0])}^t\mathbf{[V^{1/2}_{+}\ 0]}\mathbf{U}^t\boldsymbol{b}=\boldsymbol{b}^t\mathbf{S}\boldsymbol{b}\;.
\end{equation}
Thus, using the penalty matrix $\mathbf{S}$ for $\boldsymbol{b}$ is equivalent to using the identity penalty matrix for $\widetilde{\boldsymbol{b}}_r$ with smoothing parameter $\sigma^2_b$ and  no penalty on $\widetilde{\boldsymbol{b}}_f$.

To make the linear predictor $\boldsymbol{\eta}$ consistent with the previous definition, we also need to transform the design matrix $\mathbf{X}$ to $\widetilde{\mathbf{X}}$ such that $\widetilde{\mathbf{X}}^t\boldsymbol{\widetilde{b}}=\mathbf{X}^t\boldsymbol{b}$. If $\widetilde{\mathbf{V}}^{-1/2}$ is  the inverse of $\widetilde{\mathbf{V}}^{1/2}$ and  
\begin{equation}\label{eq:base_trans}\widetilde{\mathbf{X}}=\widetilde{\mathbf{V}}^{-1/2}\mathbf{U}^t\mathbf{X},
\end{equation}
then $\widetilde{\mathbf{X}}^t\boldsymbol{\widetilde{b}}=\mathbf{XU\widetilde{\mathbf{V}}^{-1/2}\widetilde{\mathbf{V}}^{1/2}}\mathbf{U}^t\boldsymbol{b}=\mathbf{X}^t\boldsymbol{b}$. We further partition $\widetilde{\mathbf{X}}$ into $\widetilde{\mathbf{X}}^t=[\widetilde{\mathbf{X}}^t_r|\widetilde{\mathbf{X}}^t_f]$, where $\widetilde{\mathbf{X}}^t_r$ are the first $K_0$ columns of $\widetilde{\mathbf{X}}^t$. With this notation the full Bayesian SoFR model where the functional parameter $\beta(t)$ is modeled nonparametrically using penalized splines is 
\begin{equation}\label{eq:stansof}
    \begin{cases}
      \qquad \mathbf{Y} \sim {\rm Exponential\_Family}(\boldsymbol{\eta},a)\;;\\
      \qquad \boldsymbol{\eta} = \eta_0\boldsymbol{J}_n+ \widetilde{\mathbf{X}}_r^t\boldsymbol{\widetilde{b}}_r+\widetilde{\mathbf{X}}_f^t\boldsymbol{\widetilde{b}}_f+\mathbf{Z}^t\boldsymbol{\gamma}\;;\\
      \qquad \boldsymbol{\widetilde{b}}_r \sim N(\boldsymbol{0},\sigma^2_b\mathbf{I})\;;\\\qquad \eta_0 \sim p(\eta_0);  \boldsymbol{\widetilde{b}}_f \sim p(\boldsymbol{\widetilde{b}}_f); \boldsymbol{\gamma}\sim p(\boldsymbol{\gamma})\;;\\
      \qquad \sigma^2_b \sim p(\sigma^2_b); a \sim p(a)\;.
    \end{cases}       
\end{equation}
where $\boldsymbol{\eta}$ is the linear predictor and $a$ is the dispersion parameter for the exponential family distribution.
Here $p(\cdot)$ is a general notation for uninformative priors on fixed effects. For example, $p(\boldsymbol{\widetilde{b}}_2)$ are  $K-K_0$ independent zero-mean normal priors with large variance (small precision) and $p(\sigma^2_b)$ is an inverse Gamma prior $IG(0.001,0.001)$ \citep{crainiceanuwinbugs,crainiceanu2010bayesian}, though other non-informative variance priors could also be used. \color{black} An alternative direct penalization approach was recently introduced by Sartini et al.\cite{sartini2025}, who proposed a fully Bayesian functional principal components analysis that projects eigenfunctions onto an orthonormal spline basis and efficiently samples the low-dimensional coefficient matrix via polar decomposition, with order constraints imposed during the sampling process to ensure identifiability.  However, here we use the reparameterization described above. \color{black}

The main contribution here is to transform the conceptual functional regression model~\eqref{eq:linearpred} into the relatively simple Bayesian model~\eqref{eq:stansof} that can be easily implemented in \texttt{Stan}. This also provides the infrastructure for adding additional smoothing and mixed effects components that would be difficult to incorporate without this initial stepping stone.


\subsection{Bayesian SoFR Implementation in Stan}\label{sec:2.3}

We now provide a detailed, step-by-step tutorial on how to implement the Bayesian SoFR model~\eqref{eq:stansof} using the \texttt{R} interface to \texttt{Stan}. We will introduce the data preparation, \texttt{Stan} program, and how to organize the results for statistical inference. Our associated \texttt{R} package \texttt{refundBayes} makes this process invisible to users, but we describe the underlying infrastructure for reasons related to reproducibility, transparency, and future software development. \color{black} We provide the detailed code in the supplementary document. The \texttt{refundBayes} package relies on the \texttt{rstan} package for Bayesian computation. Please ensure that \texttt{rstan} is installed and loaded before using functions from \texttt{refundBayes}. \color{black}

\subsubsection{Data preparation}\label{subsubsec:data}
Assume that the functional data
$\{W_i(t_m), i=1,...,n, m=1,...,M\}$ is stored in \texttt{R} as an $n\times M$ matrix \texttt{data\$wmat}, where the $i$-th row contains $M$ observations from subject $i$. To construct the matrix $\mathbf{X}$ defined in \eqref{eq:xmat}, we store the time points in an $n\times M$ matrix \texttt{data\$tmat}, where the $(i,m)$-th entry is equal to $t_{im} = t_m$, as defined in Section \ref{sec:2.1}. Additionally, let \texttt{data\$lmat} be an $n\times M$ matrix whose $(i,m)$-th entry equals $L_m=t_{m+1}-t_m$. The outcome is stored as a vector \texttt{data\$Y}. As shown in Crainiceanu et al. \cite{crainiceanu2024functional}, the frequentist SoFR can be fitted using the \texttt{mgcv::gam} function as follows:

\vspace{2mm}

\noindent\colorbox{backcolour}{
\begin{minipage}{\textwidth}
{\ttfamily\footnotesize
\comcolor{// Fit model using frequentist approaches: mgcv}

fit = mgcv::\funcol{gam}(Y $\sim$ \funcol{s}(tmat, by=lmat\opcol{*}wmat, bs=\strcol{"cr"}, k=10), 

\qquad data = data)

}
\end{minipage}
}

\vspace{2mm}

The argument \texttt{s(tmat, by=lmat*wmat, bs="cr", k=10)} constructs the matrix $\mathbf{X}$ using cubic regression splines (specified in the argument \texttt{bs="cr"}) with a $10$-dimensional basis (specified in the argument \texttt{k=10}). We will leverage this argument to construct the input for \texttt{Stan}. Specifically, calculating the matrices $\widetilde{\mathbf{X}}_r$ and $\widetilde{\mathbf{X}}_f$ in \eqref{eq:stansof} can be done directly using the \texttt{smoothCon} function in the \texttt{mgcv} package:

\vspace{2mm}

\noindent\colorbox{backcolour}{
\begin{minipage}{\textwidth}
{\ttfamily\footnotesize
\comcolor{// mgcv construction of design matrix for functional predictors}\\
smcon = mgcv::\funcol{smoothCon}(\funcol{s}(tmat, by=lmat\opcol{*}wmat, bs=\strcol{"cr"}, k=10), 

\qquad
data = data, absorb.cons = TRUE, 

\vspace{-1.5mm}
\qquad diagonal.penalty = TRUE)
}
\end{minipage}
}

\vspace{2mm}

The \texttt{smoothCon} function first creates the spline basis according to the user-supplied spline information, including the spline type and the maximum number of degrees of freedom. It then integrates the basis with the functional covariate $\mathbf{W}$ as described in \eqref{eq:linearpredmat}. By setting the argument \texttt{diagonal.penalty = TRUE}, the program will perform basis transformation introduced in \eqref{eq:base_trans} and generate the design matrix with a diagonal penalty term. \color{black} Here we set \texttt{absorb.cons = TRUE} so that the identifiability constraints are absorbed when constructing the spline basis. \color{black} The matrices $\widetilde{\mathbf{X}}_r$ and $\widetilde{\mathbf{X}}_f$ can then be obtained using the \texttt{mgcv::smooth2random} function as follows:

\vspace{2mm}

\noindent\colorbox{backcolour}{
\begin{minipage}{\textwidth}
{ \ttfamily\footnotesize
\comcolor{// Extract the transformed design matrices}\\
randeff = mgcv::\funcol{smooth2random}(smcon[[1]], name=names(data), type = 2)

X\_mat\_r = \opcol{t}(randeff\opcol{\$}rand\opcol{\$}Xr)

\vspace{-1.5mm}
X\_mat\_f = \opcol{t}(randeff\opcol{\$}rand\opcol{\$}Xf)
}
\end{minipage}
}\\

In this section we took advantage of the built-in structure of the \texttt{mgcv} package. Users may prefer to build these components themselves, which can be done following the steps described in Section~\ref{sec:2.2}.

\subsubsection{Stan code}\label{subsubsecLStan}
The transformed design matrices \color{black}$\widetilde{\mathbf{X}}_r$ and $\widetilde{\mathbf{X}}_f$ \color{black} are stored in the \texttt{data} list as \texttt{data\$X\_mat\_r} and \texttt{data\$X\_mat\_f}, respectively, in addition to the outcome $\mathbf{Y}$ stored in \texttt{data\$Y} and the design matrix for scalar predictors $\mathbf{Z}$ stored in \texttt{data\$Z}.
A typical \texttt{Stan} code consists of three main blocks: the \texttt{data} block, which defines the format of the input data; the \texttt{parameters} block, which specifies the parameters to be sampled; and the \texttt{model} block, which formulates the log-likelihood of the Bayesian model. The following \texttt{data} block defines the data components for the SoFR model:

\vspace{2mm}

\noindent\colorbox{backcolour}{
\begin{minipage}{\textwidth}
{ \ttfamily\footnotesize
\comcolor{// Stan code data block}\\
\funcol{data} \{

\qquad    \funcol{int}<lower=1> N\_num;  \comcolor{ // Total number of subjects}
    
\qquad    \funcol{int} Y[N\_num];  \comcolor{// Outcome variable}
    
\qquad    \funcol{int}<lower=1> K\_num;  \comcolor{// Number of scalar predictors}

\qquad    \comcolor{// Design matrix for the scalar predictors}

\qquad    \funcol{matrix}[K\_num, N\_num] Z\_mat;  
    
    
    
\qquad    \funcol{int} Kf;  \comcolor{// Row number of the fixed effects design matrix}
    
\qquad    \funcol{int} Kr;  \comcolor{// Row number of the random effects design matrix}
    
\qquad    \funcol{matrix}[Kf, N\_num] X\_mat\_f;  \comcolor{// Fixed effects design matrix}
    
\qquad    \funcol{matrix}[Kr, N\_num] X\_mat\_r;  \comcolor{// Random effects design matrix}

\vspace{-1.5mm}
\}
}
\end{minipage}
}

\vspace{2mm}

\noindent The parameters \texttt{N\_num}, \texttt{K\_num}, \texttt{Kf}, \texttt{Kr} are easily obtained from the model input. 

The \texttt{parameters} block specifies the parameters to be sampled in Stan:

\vspace{2mm}

\noindent\colorbox{backcolour}{
\begin{minipage}{\textwidth}
{ \ttfamily\footnotesize
\comcolor{// Stan code parameter block}\\
\funcol{parameters} \{

\qquad    \funcol{real} sigma;  \comcolor{// Smoothing parameter}
    
\qquad    \funcol{real} eta\_0;  \comcolor{// Linear predictor intercept}
    
\qquad    \funcol{vector}[Kf] betaf;  \comcolor{// Fixed effects spline coefficients}
    
\qquad    \funcol{vector}[Kr] betar;  \comcolor{// Random effects spline coefficients}
    
\qquad    \funcol{vector}[K\_num] gamma;  \comcolor{// Coefficients for scalar predictors}

\vspace{-1.5mm}
\}
}
\end{minipage}
}\\

And finally, the \texttt{model} block specifies the joint log-likelihood, including the data likelihood and the priors for the parameters. Here we show the implementation for binary outcomes using the \texttt{bernoulli\_logit\_lpmf} function in \texttt{Stan}. However, the model can as easily be applied to other exponential family distributions by calling the corresponding \texttt{Stan} functions. The code below  follows the Bayesian model described in equation \eqref{eq:stansof},  where \texttt{target} is the token for the logarithm of the likelihood in \texttt{Stan}:

\vspace{2mm}
\noindent\colorbox{backcolour}{
\begin{minipage}{\textwidth}
{ \ttfamily\footnotesize
\comcolor{// Stan code model block}\\
\funcol{model} \{

    \qquad \comcolor{// Linear predictor} 
    
    \qquad \funcol{vector}[N\_num] eta = \funcol{rep\_vector}(0.0, N\_num);
    
    \qquad eta += eta\_0 \opcol{+} X\_mat\_f' \opcol{*} betaf \opcol{+} X\_mat\_r' \opcol{*} betar \opcol{+} Z\_mat' \opcol{*} gamma;

    \qquad \comcolor{// Log-likelihood for the binary outcome}
    
    \qquad for (n in 1:N\_num) \{

    \qquad \qquad   target += \funcol{bernoulli\_logit\_lpmf}(Y[n] | eta[n]);

    \qquad \}

    \qquad \comcolor{// Set priors for the parameters}
    

    \qquad target += \funcol{normal\_lpdf}(betar | 0, sigma);
    
    \qquad target += \funcol{inv\_gamma\_lpdf}(sigma\opcol{\textnormal{\^}}2 | 0.001, 0.001);

    \vspace{-1.5mm}
\}
}
\end{minipage}
}\\

Here we do not specify explicit priors for the parameters \texttt{gamma}, \texttt{betaf}, and \texttt{eta\_0}. In this case, \texttt{Stan} implicitly assigns a uniform prior to these parameters over their respective domains. Another option is to specify noninformative priors for these parameters in the \texttt{model} block. No difference was noticed between the two approaches.


\subsubsection{Reconstructing the estimated functional coefficients}\label{subsubsec:reconstruct}
Recall that $\boldsymbol{b}=\mathbf{U}\widetilde{\mathbf{V}}^{-1/2}\widetilde{\boldsymbol{b}}$ and $\beta(t)=\sum_{k=1}^K b_k\psi_k(t)$. For every iteration of the algorithm, we obtain a sample from the joint posterior distribution of the $K\times 1$ dimensional vector of parameters $\boldsymbol{\widetilde{b}}=\{\widetilde{b}_1,...,\widetilde{b}_K\}$. Hence, at all times $t$, simultaneously, we obtain a sample of the functional coefficient as follows
$$\boldsymbol{\widetilde{b}}\quad\rightarrow \quad\boldsymbol{b}=\mathbf{U}\widetilde{\mathbf{V}}^{-1/2}\widetilde{\boldsymbol{b}}\quad\rightarrow \quad\beta(t)=\sum_{k=1}^K b_k\psi_k(t)\;.$$

\color{black}Again, these quantities can be built from scratch, but we used the \texttt{mgcv} infrastructure to compute them and construct the spline basis. However, the transformation matrix $\widetilde{\mathbf{V}}$ used for orthogonalization and scaling is not directly provided in the output of \texttt{mgcv::smoothCon}. To map the posterior samples of $\widetilde{\boldsymbol{b}}$ back to the original spline basis and reconstruct $\beta(t)$, $\widetilde{\mathbf{V}}$ is computed from scratch. \color{black}We start by extracting the spline basis $\boldsymbol{\psi}(t)$ and the corresponding penalty matrix $\mathbf{S}$using the \texttt{mgcv::smooth.construct} function.

\vspace{2mm}

\noindent\colorbox{backcolour}{
\begin{minipage}{\textwidth}
{ \ttfamily\footnotesize
\comcolor{// Construct the spline smoother components}

\vspace{-1.5mm}
splinecons = mgcv::\funcol{smooth.construct}(object, data, NULL)
}
\end{minipage}
}\\
\vspace{-1.5mm}

Here \texttt{object} is the \texttt{mgcv::s} term that specifies the spline type, parameter, and functional predictor. For example, for the cubic regression splines with $10$ degrees of freedom the \texttt{object} term would be \texttt{s(tmat, by=lmat*wmat, bs="cr", k=10)}. The spline basis $\boldsymbol{\psi}(t)$ can be extracted as:

\vspace{2mm}

\noindent\colorbox{backcolour}{
\begin{minipage}{\textwidth}
{ \ttfamily\footnotesize
\comcolor{// Extract spline basis}

\vspace{-1.5mm}
Psi\_mat = splinecons\opcol{\$}X
}
\end{minipage}
}\\
\vspace{-1.5mm}

The corresponding penalty matrix $\mathbf{S}$ can be extracted as:

\vspace{2mm}

\noindent\colorbox{backcolour}{
\begin{minipage}{\textwidth}
{ \ttfamily\footnotesize
\comcolor{// Extract penalty matrix}

\vspace{-1.5mm}
S\_mat = splinecons\opcol{\$}S
}
\end{minipage}
}\\
\vspace{-1.5mm}

The rank, $K_0$, of the penalty matrix $\mathbf{S}$ can be extracted as:

\vspace{2mm}

\noindent\colorbox{backcolour}{
\begin{minipage}{\textwidth}
{ \ttfamily\footnotesize
\comcolor{// Extract the rank of S}

\vspace{-1.5mm}
rank = splinecons\opcol{\$}rank
}
\end{minipage}
}\\
\vspace{-1.5mm}

As $\mathbf{S}=\mathbf{UVU}^t$, the $\mathbf{U}$ and $\mathbf{V}$ matrices are obtained as:

\vspace{2mm}

\noindent\colorbox{backcolour}{
\begin{minipage}{\textwidth}
{ \ttfamily\footnotesize
\comcolor{// Spectral decomposition}\\
eigendecomp = \funcol{eigen}(S\_mat, symmetric = TRUE)

\comcolor{// Eigenfunctions}\\
U\_mat = eigendecomp\opcol{\$}vectors

\comcolor{// Eigenvalues}

\vspace{-1.5mm}
V\_vec = eigendecomp\opcol{\$}value[1:rank]
}
\end{minipage}
}\\

Now we calculate the diagonal elements of the $\widetilde{\mathbf{V}}^{1/2}$ matrix used in the \texttt{smoothCon} function:

\vspace{2mm}

\noindent\colorbox{backcolour}{
\begin{minipage}{\textwidth}
{ \ttfamily\footnotesize
V = \funcol{rep}(1, \funcol{ncol}(X\_mat))

\comcolor{// First K0 entries}\\
V[1:rank] = \funcol{sqrt}(V\_vec)

\comcolor{// Calculate the remaining K-K0 entries}\\
col.norm = \funcol{colSums}((X\_mat \opcol{\%*\%} U\_mat)\opcol{\textnormal{\^}}2)

col.norm = col.norm \opcol{/} V\opcol{\textnormal{\^}}2

av.norm = \funcol{mean}(col.norm[1:rank])

for (i in (rank \opcol{+} 1):\funcol{ncol}(X\_mat)) \{

  \qquad V[i] = \funcol{sqrt}(col.norm[i] \opcol{/} av.norm)
  
\}

}
\end{minipage}
}\\

Here \texttt{X\_mat} is the transposed of the original design matrix $\mathbf{X}$. The \texttt{smoothCon} function uses a slightly different definition for the $K\times K$ matrix $\widetilde{\mathbf{V}}$ than what is presented in \eqref{eq:transb}. Specifically, both the \texttt{smoothCon} function and equation (\ref{eq:transb}) define the first $K_0$ entries of the diagonal vector of the matrix $\widetilde{\mathbf{V}}$ as the positive eigenvalues of $\mathbf{S}$. However, unlike (\ref{eq:transb}), where the remaining $K-K_0$ diagonal entries are constant and equal to $1$, the \texttt{smoothCon} function defines them based on the norm of the design matrix $\mathbf{X}$ (as illustrated in the  code). This exact construction of $\widetilde{\mathbf{V}}$ is necessary if \texttt{mgcv} is used to construct the model components, but other options could be used, as well, as long as one keeps track of the reparametrization operators.

As $\boldsymbol{b}=\mathbf{U}\widetilde{\mathbf{V}}^{-1/2}\widetilde{\boldsymbol{b}}$, the posterior samples of the spline coefficients, $\boldsymbol{b}^q$, $q=1,\ldots,Q$, can be obtained as:

\vspace{2mm}

\noindent\colorbox{backcolour}{
\begin{minipage}{\textwidth}
{ \ttfamily\footnotesize
\comcolor{// Transform the posterior sample of the spline coefficient}

\vspace{-1.5mm}
 beta.sample.untilde = (U\_mat \opcol{\%*\%} diag(1 \opcol{/} V)) \opcol{\%*\%}  beta.sample
}
\end{minipage}
}\\
\vspace{-1.5mm}

Here \texttt{beta.sample} contains the  posterior sample of the transformed spline coefficients $\boldsymbol{\widetilde{b}}$. Having all necessary ingredients, the posterior sample of the functional effect $\beta(t)$ can be calculated at every iteration, $q=1,\ldots,Q$, as
\begin{equation}
    \beta^q(t)=\sum_{k=1}^K b_k^q\psi_{k}(t), \;\; q = 1,\ldots, Q\;.
\end{equation}
The posterior samples of the functional effect $\beta^q(t)$ are stored in a $Q\times M$ dimensional matrix, where each row corresponds to a posterior iteration and each column corresponds to a location, $t$, on the functional domain. The column mean, $\widehat{\beta}(t)$, and the covariance of this matrix, $\widehat{C}_\beta$, can then be used to provide pointwise and correlation and multiplicity adjusted (CMA) credible intervals; see \cite{crainiceanu2024functional} for an in-depth description of CMA.  Pointwise confidence intervals can also be obtained using pointwise quantiles (separately at every $t$), which can be especially useful when the posterior distribution of $\beta(t)$ is not Gaussian.
 \color{black} In the supplementary R Markdown file, we provide the complete code for our Bayesian Stan model. \color{black}

\subsection{Simulations}\label{sec:2.4}
Using simulations, we compare the inferential accuracy of the Bayesian SoFR implementation with that of the state-of-the-art frequentist approach. 

\subsubsection{Data generating mechanism} We consider the total of $n=100$, $200$, $300$, and  $500$ subjects and one functional covariate observed at $T=50$ equally-spaced points in $[0,1]$. The functional covariates $\{W_i(t_j), t_j\in [0,1]\}_{t=1}^T$ are generated using the first $4$ principal components estimated from our NHANES case study \citep{leroux2019cao}. The functional coefficient  is $\beta(t)=\big(0.084-(t-0.5)^2\big)\times\tau$, $t\in [0,1],$ with $\tau=1$, $2$, $3$, and $5$ controlling the strength of the signal. The outcome variables $Y_i, i=1,\ldots,n$ are generated from either a Gaussian distribution with mean $\eta_i = \frac{1}{T}\sum_{j=1}^T W_i(t_j)\beta(t_j)$ and standard deviation of $1.5$, or a Bernoulli distribution with success probability $\exp(\eta_i)/\{1+\exp(\eta_i)\}$. 

\subsubsection{Competing methods} We compare the performance of our Bayesian \texttt{Stan} program with the results obtained using the \texttt{gam} function from the \texttt{R} package \texttt{mgcv}. For both Bayesian and frequentist methods, we use a cubic regression spline (\texttt{bs="cr"}) basis with a maximum of $10$ degrees of freedom (\texttt{k=10}). As described in Crainiceanu et al.\cite{crainiceanu2024functional}, one can also use the \texttt{refund::pfr} \citep{goldsmith2011penalized} function in \texttt{R} to conduct the analyses.

\subsubsection{Metrics} For each simulation scenario, we calculate the relative integrated squared error (RISE) of the estimated functional effect, defined as: $\int \{\beta(t)-\widehat{\beta}(t)\}^2dt / \int \beta^2(t)dt$, where $\beta(t)$ and $\widehat{\beta}(t)$ are the true and the estimated functional coefficient, respectively. We compare the average empirical coverage rate for the pointwise $95\%$ confidence intervals. \color{black} In addition to evaluating the coefficients, we also assess the predictive performance on a new set of $500$ study participants  by comparing the predicted outcomes to the corresponding true values.  \color{black}

\subsubsection{Results} Simulation results are presented in Table \ref{tab:sof_n} (for Gaussian outcomes) and Table \ref{tab:sof_b} (for binary outcomes), based on $500$ simulations for each scenario. We report the median relative integrated squared error (RISE) and the mean coverage rate. As sample size and signal level vary, the median RISE ranges from near zero (indicating high estimation accuracy) to values above $5$ (indicating that the estimation is far from the truth). Although the estimation performance varies across scenarios, the Bayesian approach consistently demonstrates similar performance with the frequentist approach for both Gaussian and binary outcomes. Both methods exhibit similar  RISE, with the Bayesian approach achieving a slightly higher coverage rate of the pointwise confidence intervals. \color{black} The supplementary material provides the mean computation time (in minutes) for the Bayesian Stan program.\color{black}
 
\begin{center}
\begin{table*}[]
\centering
\caption{\color{black} Simulation results for scalar-on-function regression (SoFR) comparing the frequentist and Bayesian approach for Gaussian outcomes for different sample sizes, $n$,  and signal levels, $\tau$.  The median RISE, mean coverage rate of the 95\% credible/confidence intervals, and prediction accuracy are reported. \label{tab:sof_n}}
\resizebox{0.8\textwidth}{!}{
\color{black}
\begin{tabular}{ccccccccccccc}
\hline
& &
\multicolumn{3}{c}{\textbf{n=100}} & 
\multicolumn{3}{c}{\textbf{n=200}} &
\multicolumn{3}{c}{\textbf{n=300}} &
\multicolumn{2}{c}{\textbf{n=500}} \\
\cline{3-4} \cline{6-7} \cline{9-10} \cline{12-13} 
  &  & Bayes & Freq &  & Bayes & Freq &  & Bayes & Freq &  & Bayes & Freq \\ 
 $\tau = 1$ & RISE & 4.694 & 6.025 &  & 2.431 & 2.973 &  & 1.837 & 2.123 &  & 1.367 & 1.55 \\ 
   & Coverage & (96.4) & (94.5) &  & (96.7) & (94.3) &  & (97.1) & (93.8) &  & (96.3) & (92.8) \\ 
   & Prediction & 3.653 & 5.785 &  & 1.8 & 2.926 &  & 1.365 & 2.112 &  & 1.017 & 1.459 \\ 
 $\tau = 2$ & RISE & 1.525 & 1.725 &  & 0.917 & 0.925 &  & 0.623 & 0.554 &  & 0.394 & 0.352 \\ 
   & Coverage & (95.9) & (92.9) &  & (97.1) & (95.1) &  & (97.2) & (95.2) &  & (97.9) & (95.9) \\ 
   & Prediction & 1.121 & 1.827 &  & 0.691 & 0.863 &  & 0.457 & 0.555 &  & 0.278 & 0.339 \\ 
 $\tau = 3$ & RISE & 0.892 & 0.842 &  & 0.413 & 0.353 &  & 0.298 & 0.287 &  & 0.154 & 0.151 \\ 
   & Coverage & (96.5) & (93.5) &  & (97.4) & (96) &  & (98.9) & (97.6) &  & (99.5) & (98.4) \\ 
   & Prediction & 0.683 & 0.814 &  & 0.294 & 0.349 &  & 0.196 & 0.253 &  & 0.103 & 0.142 \\ 
 $\tau = 5$ & RISE & 0.334 & 0.314 &  & 0.148 & 0.148 &  & 0.111 & 0.103 &  & 0.067 & 0.063 \\ 
   & Coverage & (98.3) & (96.6) &  & (99.4) & (97.5) &  & (99.5) & (98) &  & (99.8) & (98.9) \\ 
   & Prediction & 0.239 & 0.293 &  & 0.1 & 0.132 &  & 0.069 & 0.094 &  & 0.042 & 0.056 \\ 
    \hline
    
  \end{tabular}

  }
\end{table*}

\end{center}

\begin{center}
\begin{table*}[]
\centering
\caption{\color{black} Simulation results for Bernoulli outcomes using the same structure and measures as Table~\ref{tab:sof_n}.  \label{tab:sof_b}}
\resizebox{0.8\textwidth}{!}{
\color{black}
\begin{tabular}{ccccccccccccc}
\hline
& &
\multicolumn{3}{c}{\textbf{n=100}} & 
\multicolumn{3}{c}{\textbf{n=200}} &
\multicolumn{3}{c}{\textbf{n=300}} &
\multicolumn{2}{c}{\textbf{n=500}} \\
\cline{3-4} \cline{6-7} \cline{9-10} \cline{12-13} 
  &  & Bayes & Freq &  & Bayes & Freq &  & Bayes & Freq &  & Bayes & Freq \\ 
$\tau = 1$ & RISE & 9.792 & 11.087 &  & 4.529 & 5.577 &  & 3.117 & 3.611 &  & 2.081 & 2.311 \\ 
   & Coverage & (96.4) & (95.7) &  & (96) & (94.3) &  & (96.8) & (94.7) &  & (96.3) & (93.5) \\ 
   & Prediction & 6.618 & 10.737 &  & 3.101 & 5.497 &  & 2.141 & 3.486 &  & 1.499 & 2.357 \\ 
 $\tau = 2$ & RISE & 2.722 & 2.932 &  & 1.525 & 1.687 &  & 1.031 & 1.1 &  & 0.71 & 0.634 \\ 
   & Coverage & (95.6) & (94.3) &  & (95.8) & (93.1) &  & (96.3) & (94) &  & (97.7) & (96) \\ 
   & Prediction & 1.862 & 2.925 &  & 1.155 & 1.607 &  & 0.765 & 1.055 &  & 0.514 & 0.633 \\ 
 $\tau = 3$ & RISE & 1.425 & 1.525 &  & 0.829 & 0.782 &  & 0.604 & 0.521 &  & 0.324 & 0.287 \\ 
   & Coverage & (96.3) & (93.7) &  & (96.7) & (94.3) &  & (97.5) & (95.6) &  & (98.6) & (97.2) \\ 
   & Prediction & 1.084 & 1.47 &  & 0.626 & 0.721 &  & 0.417 & 0.473 &  & 0.214 & 0.279 \\ 
 $\tau = 5$ & RISE & 0.748 & 0.667 &  & 0.34 & 0.313 &  & 0.213 & 0.198 &  & 0.126 & 0.117 \\ 
   & Coverage & (97.2) & (95.9) &  & (98.5) & (97) &  & (99) & (97.9) &  & (99.3) & (97.8) \\ 
   & Prediction & 0.539 & 0.61 &  & 0.234 & 0.264 &  & 0.138 & 0.19 &  & 0.087 & 0.108 \\ 
    \hline
    
  \end{tabular}

  }
\end{table*}
\end{center}

\section{Bayesian Functional Cox Regression}\label{sec:3}

In this section, we describe the process of fitting a Bayesian Functional Cox Regression (FCR) model with a time-to-event outcome in \texttt{Stan}. Surprisingly, there are few published frequentist approaches for analyzing such data; see, for example, Gellar et al.\cite{gellar2015cox}; Kong et al.\cite{kong2018flcrm}; Qu et al.\cite{qu2016optimal}, who proposed different versions of the “linear functional Cox model.” Recently, Cui et al.\cite{cui2021additive} introduced the Additive Functional Cox Model (AFCM), which extended the methods introduced by  Gellar et al. \cite{gellar2015cox} to account for non-linear functional effects, as introduced by McLean et al. \cite{mclean2014functional} for generalized functional regression models. Extending these methods to Bayesian analysis is not straightforward, as a full likelihood needs to be specified, including modeling of the baseline hazard function. In this tutorial, we  model the baseline hazard function using splines \citep{brilleman2020bayesian,Cheng01092009}, and we have adopted a new combination of priors to ensure a robust performance of the \texttt{Stan} implementation. 

\subsection{The Functional Cox Regression Model}\label{subsec:FCRM}
For subject $i=1,...,n$, denote by $T_i$ the event time and by $C_i$ the censoring time, where $T_i$ is observed only when $T_i\leq C_i$. The observed data is $[Y_i, \delta_i, \mathbf{Z}_i, \{W_i(t_{im}), t_{im}\in [0,1]\}]$, where $Y_i = \min(T_i, C_i)$, and $\delta_i$ is the indicator of right censoring with $\delta_i=0$ if $Y_i=T_i$ (an observed event), and $\delta_i=1$ if $Y_i<T_i$ (a censored event), $\mathbf{Z}_i$ is the $p\times 1$ dimensional vector of scalar predictors, and $\{W_i(t_{im}), t_{im}\in [0,1]\}$ for $m=1,\ldots,M_i$ is the functional predictor. 

If $T$ is a random variable, its hazard function is defined as the instantaneous rate of occurrence for the event at time $t$:
\begin{equation}
    h_T(t)=\lim_{\Delta t\to 0} \frac{Pr(t\leq T\leq t+\Delta t|T>t)}{\Delta t}\;.
\end{equation}
The cumulative hazard function is defined as  $H_T(t)=\int_{u=0}^t h(u)du$, the cumulative distribution function (cdf) is defined as $F_T(t)=P(T\leq t)$, and the survival function is defined as $S_T(t)=1-F_T(t)$. It can easily be shown that $S_T(t)=\exp\{-H(t)\}$. To model the individual hazard  function we consider a natural extension of the Cox proportional hazards model \citep{Cox1972}, which assumes that $h_i(t)=h_0(t)\exp(\eta_i)$, where $h_0(t)$ is the baseline hazard function, and $\eta_i$ is the linear predictor for subject $i$ defined in equation \eqref{eq:theta}. The log-likelihood for this model has the following form
\begin{equation}\label{eq:coxlikelihood}
        \begin{split}
            l(\boldsymbol{Y, \delta};h_0,\boldsymbol{\eta}) = \sum_{i=1}^n \bigg[(1-\delta_i)[\log\{h_0(y_i)\}+\eta_i\\
            \qquad -H_0(y_i)\exp(\eta_i)]+\delta_i\{-H_0(y_i)\exp(\eta_i)\}\bigg]\;,\\
        \end{split}
\end{equation}
where $H_0(t)=\int_{u=0}^t h_0(u)du$ is the cumulative baseline hazard function.

\subsection{Model of the Hazard Function using M-splines}\label{sec:3.2}
Similar to  Brilleman et al. \citep{brilleman2020bayesian}, we model the baseline hazard function $h_0(t)$ using an M-spline basis \citep{ramsay1988}, such that $h_0(t)=\sum_{l=1}^L c_lM_l(t;\boldsymbol{k},\tau)$,
where $M_l(t;\boldsymbol{k},\tau)$ denotes the $l$-th M-spline basis with knots $\boldsymbol{k}$ and degree of freedom $\tau$, and $\boldsymbol{c}=(c_1,...,c_L)^t$ are the spline coefficients. We require that $c_l\geq 0$ and $\sum_{l=1}^Lc_l>0$, which ensures that $h_0(t)\geq 0$ and $h_0(t)\neq 0$ (events do occur).
\color{black}
M-splines are a family of non-negative, piecewise polynomial basis functions integrate to one over their support. To model the cumulative baseline hazard function, we use I-splines, which are the integrated forms of M-splines and therefore yield monotone functions.
\color{black}
If $I_l(t;\boldsymbol{k},\tau)=\int_0^t M_l(u;\boldsymbol{k},\tau)du$ denotes the corresponding I-spline basis function \citep{ramsay1988}, the cumulative baseline hazard function 
$H_0(t)=\sum_{l=1}^L c_lI_l(t;\boldsymbol{k},\tau)$ is non-decreasing by construction.

The I-spline coefficients, $\boldsymbol{c}$, and the intercept, $\eta_0$, in the linear predictor $\eta_i$ are not simultaneously identifiable. Indeed, for any $a > 0$ the parameters $(\boldsymbol{c}, \eta_0)$ and $(\widetilde{\boldsymbol{c}}, \widetilde{\eta}_0)=\{a\boldsymbol{c}, \eta_0-\log(a)\}$ correspond to the same hazard $h_0(t)\exp(\eta_i)$.  To make the model identifiable, we impose the constraint $\sum_{l=1}^L c_l=1$ by assuming that the coefficients $\boldsymbol{c}$ have a non-informative Dirichlet prior $D(\boldsymbol{c};\boldsymbol{\alpha})$, where $\boldsymbol{\alpha}=(1,...,1)$.  The full Bayesian functional Cox regression model is 
\begin{equation}\label{eq:stanfunc}
    \begin{cases}
      \qquad \boldsymbol{Y} \sim l(\boldsymbol{Y},\boldsymbol{\delta};h_0,\boldsymbol{\eta})\;;\\
      \qquad \boldsymbol{\eta} = \eta_0\boldsymbol{J}_n+ \widetilde{\mathbf{X}}_r^t\boldsymbol{\widetilde{b}}_1+\widetilde{\mathbf{X}}_f^t\boldsymbol{\widetilde{b}}_2+\mathbf{Z}^t\boldsymbol{\gamma}\;;\\
      \qquad \boldsymbol{\widetilde{b}}_1 \sim N(\boldsymbol{0},\sigma^2_b\mathbf{I})\;;\\
      \qquad \eta_0 \sim p(\eta_0);  \boldsymbol{\widetilde{b}}_2 \sim p(\boldsymbol{\widetilde{b}}_2); \boldsymbol{\gamma}\sim p(\boldsymbol{\gamma}); \sigma^2_b \sim p(\sigma^2_b)\;;\\
      \qquad h_0(t)=\sum_{l=1}^L c_lM_l(t;\boldsymbol{k},\tau)\;;\\
      \qquad \boldsymbol{c}\sim D(\boldsymbol{c};\boldsymbol{\alpha})\;,
    \end{cases}       
\end{equation}
where most components are similar to model \eqref{eq:stansof}, except the likelihood (first line of the model) and the specification of the baseline hazard using I-splines (last two lines of the model).

\subsection{Bayesian FCR Implementation in Stan}\label{subsec:RforCoxSoFR}

\subsubsection{Data preparation}

Suppose that the the data $\boldsymbol{Y}$ are stored as a vector \texttt{y} in \texttt{R}, and the event indicator $1-\boldsymbol{\delta}$ is stored in \texttt{data\$event}.
We first obtain the M-spline basis $\boldsymbol{M}=\{M_1(t;\boldsymbol{k},\tau),...,M_L(t;\boldsymbol{k},\tau)\}$ using, for example, the \texttt{splines2::mSpline} \citep{splines2-paper,splines2-package} function in \texttt{R}.

\vspace{2mm}

\noindent\colorbox{backcolour}{
\begin{minipage}{\textwidth}
{ \ttfamily\footnotesize	 
\comcolor{// Construct the M-spline basis}\\
Mbasis = splines2::\funcol{mSpline}(y, Boundary.knots=\funcol{c}(min.bound, max.bound), df=5, intercept=TRUE)

}
\end{minipage}
}\\

The lower bound of the knots is set to be slightly less than the minimal value of $\boldsymbol{Y}$, and the upper bound is set to be slightly larger than the maximum value of $\boldsymbol{Y}$. Here, we choose the degrees of freedom of the spline basis to be $L=5$ by setting \texttt{df=5}. The other parameters are automatically determined by the \texttt{mSpline} function. The corresponding I-spline basis can be obtained similarly by calling the \texttt{iSpline} function.

\vspace{2mm}

\noindent\colorbox{backcolour}{
\begin{minipage}{\textwidth}
{ \ttfamily\footnotesize	 
\comcolor{// Construct the I-spline basis}\\
Ibasis = splines2::\funcol{iSpline}(y, Boundary.knots=\funcol{c}(min.bound,  max.bound), df=5)

}
\end{minipage}
}\\

\vspace{-4.5mm}
\subsubsection{Stan code}

The \texttt{Stan} program for the linear functional Cox model is shown below. Specifying the log-likelihood requires both the baseline hazard rate and the linear predictor. Therefore, we need to add the censoring indicator variable, the M-spline basis, and the I-spline basis to the \texttt{data} block, in addition to those defined in Section \ref{sec:2.3}:

\vspace{2mm}

\noindent\colorbox{backcolour}{
\begin{minipage}{\textwidth}
{ \ttfamily\footnotesize	 
\comcolor{// Stan code data block}\\
\funcol{data}\{ 

\qquad    ......
    
\qquad    \funcol{array}[N\_num] int cens;
    
\qquad    \funcol{int} L\_num;
    
\qquad    \funcol{matrix}[N\_num, L\_num] Mbasis;
    
\qquad    \funcol{matrix}[N\_num, L\_num] Ibasis;

\vspace{-1.5mm}
\}
}
\end{minipage}
}\\

\vspace{2mm}

Here \texttt{cens} is the censoring indicator $\delta_i, i=1,...,n$, and \texttt{L\_num} is the number of M-spline basis, $L$. \texttt{Mbasis} is an $n \times L$ dimensional matrix, with the $(i,l)$-th entry equal to $M_l(Y_i;\boldsymbol{k},\tau)$; similarly \texttt{Ibasis} is the $n \times L$ dimensional matrix corresponding to the integrated basis.

The \texttt{parameter} block, in addition to those defined in Section \ref{sec:2.3}, includes the basis coefficients for the M-splines and I-splines, $\boldsymbol{c}$:

\vspace{2mm}

\noindent\colorbox{backcolour}{
\begin{minipage}{\textwidth}
{ \ttfamily\footnotesize	 
\comcolor{// Stan code parameter block}\\
\funcol{parameters}\{ 

\qquad    ......
    
\qquad    \funcol{simplex}[L\_num] c;

\vspace{-1.5mm}
\}
}
\end{minipage}
}\\

\color{black}The spline coefficients $\mathbf{c}$ are defined as a simplex variable because its entries are non-negative and sum to one. \color{black}
The baseline hazard and the cumulated baseline hazard function share the same parameters and they can be calculated as \texttt{Mbasis*c} and \texttt{Ibasis*c}, respectively. 

The \texttt{model} block defines several functions required for calculating the log-likelihood in \eqref{eq:coxlikelihood}.

\vspace{2mm}

\noindent\colorbox{backcolour}{
\begin{minipage}{\textwidth}
{ \ttfamily\footnotesize	 
\comcolor{// Stan code function block}\\
\comcolor{// eta: linear predictor}\\ 
\comcolor{// bhaz and cbhaz: baseline and cumulative baseline hazard}\\
\funcol{functions} \{

\qquad  real \funcol{cox\_log\_lhaz}(real y, real eta, real bhaz, real cbhaz) 

\qquad \qquad    \{return \funcol{log}(bhaz) \opcol{+} eta;\}
 
\qquad  real \funcol{cox\_log\_lccdf}(real y, real eta, real bhaz, real cbhaz) 

\qquad \qquad    \{return - cbhaz \opcol{*} exp(eta);\}

\qquad  real \funcol{cox\_log\_lpdf}(real y, real eta, real bhaz, real cbhaz) 

\qquad \qquad   \{return \funcol{cox\_log\_lhaz}(y, eta, bhaz, cbhaz) \opcol{+}

\qquad \qquad \qquad \qquad \funcol{cox\_log\_lccdf}(y | eta, bhaz, cbhaz);\}

\vspace{-1.5mm}
\}
}
\end{minipage}
}\\

The code for the \texttt{model} block is shown below, where the log-likelihood is calculated using the functions defined in the \texttt{functions} block:

\vspace{2mm}

\noindent\colorbox{backcolour}{
\begin{minipage}{\textwidth}
{ \ttfamily\footnotesize	 
\comcolor{// Stan code model block}\\
\funcol{model} \{

\qquad    \comcolor{// Construct the baseline hazard rate and cumulated baseline hazard rate}
    
\qquad    \funcol{vector}[N\_num] bhaz = Mbasis \opcol{*} c;
    
\qquad    \funcol{vector}[N\_num] cbhaz = Ibasis \opcol{*} c;

\qquad    \comcolor{// Linear predictor }
    
\qquad    \funcol{vector}[N\_num] eta = \funcol{rep\_vector}(0.0, N\_num);
    
\qquad    eta += eta\_0 \opcol{+} X\_mat\_f' \opcol{*} betaf \opcol{+} X\_mat\_r' \opcol{*} betar \opcol{+} Z\_mat' \opcol{*} gamma;

\qquad    \comcolor{// Log-likelihood for the time-to-event outcome}
    
\qquad    for (n in 1:N\_num) \{

\qquad\qquad      if (cens[n] == 0) \{

\qquad\qquad \qquad target += \funcol{cox\_log\_lpdf}(Y[n] | eta[n], bhaz[n], cbhaz[n]);

\qquad\qquad \}
        
\qquad \qquad      else if (cens[n] == 1) \{

\qquad\qquad\qquad target += \funcol{cox\_log\_lccdf}(Y[n] | eta[n], bhaz[n], cbhaz[n]);

\qquad\qquad \}

\qquad \}

\qquad    \comcolor{// Set priors for the parameters}

\qquad    \funcol{vector}[L\_num] alpha = \funcol{rep\_vector}(1, L\_num);
    
\qquad    target += \funcol{dirichlet\_lpdf}(c | alpha);
    
\qquad    target += \funcol{normal\_lpdf}(betar | 0, sigma);
    
\qquad    target += \funcol{inv\_gamma\_lpdf}(sigma\opcol{\textnormal{\^}}2 | 0.001, 0.001);

\vspace{-1.5mm}
\}
}
\end{minipage}
}\\

As in Section \ref{sec:2}, we omit an explicit prior expression for the parameters \texttt{gamma}, \texttt{betaf}, and \texttt{eta\_0}, which are then set automatically by \texttt{Stan}.


\vspace{-5.5mm}
\subsection{Simulations}\label{sec:3.4}
Similar to SoFR, we compare the performance of our Bayesian Functional Cox Regression model implemented in \texttt{Stan} with the frequentist implementation in the \texttt{mgcv::gam} function. \color{black} The \texttt{mgcv} package handles Cox proportional hazards models through its \texttt{cox.ph()} family, which fits the model using partial likelihood and a Breslow-type nonparametric estimator of the baseline hazard function. \color{black}

\subsubsection{Data generating mechanism}
We use a simulation procedure similar to that introduced in Section~\ref{sec:2.4}. Following Cui et al.\cite{cui2021additive}, we generate the survival outcome based on our case study NHANES data. More specifically, we generate the baseline hazard function according to the estimated baseline hazard function from NHANES. The hazard function for each subject can then be calculated using the generated baseline hazard function and the simulated linear predictors. Once the hazard function is calculated,  event times are simulated according to the one-to-one relationship between the hazard function and the distribution function. The censoring time for each subject is randomly sampled among the simulated event times; see the supplementary materials for a complete implementation of the simulations. 

\subsubsection{Results}
The simulation results are presented in Table \ref{tab:funcox}. Consistent with the results for SoFR, our proposed Bayesian algorithm has a good performance in terms of the relative integrated squared error (RISE) and coverage rate compared with the frequentist method. 

\begin{center}
\begin{table*}[]
\centering
\caption{\color{black} Simulation results for functional Cox regression comparing the frequentist and Bayesian approach for time-to-event outcomes for different sample sizes, $n$,  and signal levels, $\tau$.  The median RISE, mean coverage rate of the 95\% credible/confidence intervals, and prediction accuracy are reported.  \label{tab:funcox}}
\resizebox{0.8\textwidth}{!}{
\color{black}
\begin{tabular}{ccccccccccccc}
\hline

& &
\multicolumn{3}{c}{\textbf{n=100}} & 
\multicolumn{3}{c}{\textbf{n=200}} &
\multicolumn{3}{c}{\textbf{n=300}} &
\multicolumn{2}{c}{\textbf{n=500}} \\
\cline{3-4} \cline{6-7} \cline{9-10} \cline{12-13} 
  &  & Bayes & Freq &  & Bayes & Freq &  & Bayes & Freq &  & Bayes & Freq \\ 
  \hline
 $\tau = 1$ & RISE & 4.281 & 5.053 &  & 2.288 & 2.52 &  & 1.763 & 1.838 &  & 1.261 & 1.247 \\ 
   & Coverage & (96.3) & (96.2) &  & (96.9) & (95.7) &  & (96.1) & (95.3) &  & (96.2) & (95.6) \\ 
   & Prediction & 3.095 & 3.29 &  & 1.637 & 1.874 &  & 1.284 & 1.366 &  & 0.9 & 0.981 \\ 
 $\tau = 2$ & RISE & 1.626 & 1.728 &  & 0.814 & 0.801 &  & 0.601 & 0.519 &  & 0.362 & 0.297 \\ 
   & Coverage & (94.8) & (94.5) &  & (96.7) & (96.9) &  & (97.3) & (97.8) &  & (98.4) & (98.6) \\ 
   & Prediction & 1.169 & 1.303 &  & 0.613 & 0.557 &  & 0.413 & 0.351 &  & 0.241 & 0.213 \\ 
 $\tau = 3$ & RISE & 0.852 & 0.841 &  & 0.417 & 0.359 &  & 0.24 & 0.201 &  & 0.146 & 0.126 \\ 
   & Coverage & (95.9) & (95.9) &  & (97.9) & (98.3) &  & (98.7) & (99.1) &  & (99.2) & (99.4) \\ 
   & Prediction & 0.664 & 0.578 &  & 0.283 & 0.247 &  & 0.164 & 0.148 &  & 0.095 & 0.09 \\ 
 $\tau = 5$ & RISE & 0.3 & 0.276 &  & 0.146 & 0.128 &  & 0.099 & 0.092 &  & 0.061 & 0.058 \\ 
   & Coverage & (98.2) & (98.9) &  & (99) & (99.5) &  & (99.4) & (99.5) &  & (99.7) & (99.6) \\ 
   & Prediction & 0.207 & 0.196 &  & 0.095 & 0.089 &  & 0.063 & 0.059 &  & 0.041 & 0.039 \\ 
    \hline
    
  \end{tabular}

  }
\end{table*}
\end{center}


\section{Bayesian Joint Functional Models with FPCA}\label{sec:4}

In the models presented in Sections \ref{sec:2} and \ref{sec:3}, the observed data are directly used as functional predictors in the regression framework. While this approach is widely adopted in the functional data analysis literature, the observed functional data are often measured with error and/or not measured at the same locations. A common strategy to address these challenges is to first apply functional principal component analysis (FPCA) to the observed data, obtain the predicted values on a grid, and use these predictions as functional predictors in the regression model. This  two-step approach treats the FPCA predictions as fixed inputs to the regression, ignoring the uncertainty of the score estimators, which could affect estimation accuracy and coverage of the confidence intervals. To address these potential issues, we adopt a joint modeling approach that simultaneously estimates the PC scores and fits the regression model \citep{crainiceanu2009generalized}. The approaches are similar for exponential family and time-to-event outcomes. Therefore, we only describe how to do this for time-to-event outcomes. 

\subsection{The Joint Functional Models with FPCA}\label{subsec:Bayesian_ME}
\subsubsection{FPCA model and notation}

Assume that the observed functional covariate ${W}_i(t)$ has the following structure
${W}_i(t)=\mu(t)+D_i(t)+\epsilon_i(t)$,
where $D_i(t)$ is a mean $0$ stochastic process with covariance operator $K_D(t,s)=\textnormal{cov}\{D_i(t),D_i(s)\}$, and $\epsilon_i(t)$ is a white noise process. For simplicity, we assume that $\mu(t)=0$ and by Kosambi–Karhunen–Loève (KKL) theorem \citep{karhunen1947under}, $D_i(t)=\sum_{j=1}^\infty \xi_{ij}\phi_j(t)$, where $\phi_j(t)$ are orthonormal eigenfunctions and $\xi_{ij}$ are mutually uncorrelated random variables with eigenvalues $\lambda_j$, respectively, where $\lambda_j\geq 0$ is a decreasing sequence. Assuming that the first $J$ eigenfunctions provide a good approximation for $D_i(t)$, it follows that $W_i(t)\approx\sum_{j=1}^J \xi_{ij}\phi_j(t) + \epsilon_i(t)$. A standard simplifying assumption is that $\xi_{ij} \sim N(0,\lambda_j)$ and $\epsilon_i(t)\sim N(0,\sigma_\epsilon^2)$. 

\subsubsection{The joint functional model}
The joint model considers the regression of outcomes on the latent trajectory $D_i(t)$ and not on the observed functions $W_i(t)$. The difference is that $W_i(t)$ is observed with noise (sometimes referred to as ``measurement error'' for functional data), which could influence the point estimators and confidence intervals. To address this problem we consider models of the type
\begin{equation}
    \begin{cases}\label{eq:jointmodel}
       \qquad Y_i\sim l(Y_i;\eta_i)\;;\\
       \qquad \eta_i=\eta_0+\int_0^1 D_i(t)\beta(t)dt + \mathbf{Z}^T_i\boldsymbol{\gamma}\;;\\
       \qquad W_i(t) = D_i(t)+\epsilon_i(t)\;,
    \end{cases}       
\end{equation}
where $l(Y_i;\eta_i)$ is the conditional distribution of the outcome $Y_i$ given the linear predictor $\eta_i$, and could correspond to either the exponential or the time-to-event families of distributions. Other conditional distributions could also be considered, but are not addressed in this tutorial.

As $\beta(t)=\sum_{k=1}^Kb_k\psi_k(t)$ and $D_i(t)=\sum_{j=1}^J \xi_{ij}\phi_j(t)$, the linear predictor $\eta_i$ can be expressed as 
 
\begin{equation}\label{eq:fpcapred}
    \begin{split}
        \eta_i&=\eta_0+\sum_{j=1}^J\sum_{k=1}^K \xi_{ij}b_k\int_0^1 \phi_j(t)\psi_k(t)dt + \mathbf{Z}^T_i\boldsymbol{\gamma}\\
    &=\eta_0+\sum_{j=1}^J\sum_{k=1}^K \xi_{ij}b_k X_{jk} + \mathbf{Z}^T_i\boldsymbol{\gamma}\\
        &=\eta_0+ \boldsymbol{\xi}_i^t \mathbf{X}^t\boldsymbol{b}+\mathbf{Z}_i^t\boldsymbol{\gamma}\;,
    \end{split}
\end{equation}
where $X_{jk}=\int_0^1 \phi_j(t)\psi_k(t)dt$ is the inner product of the eigenfunction $\phi_j(t)$ and the spline basis $\psi_k(t)$, and $\mathbf{X}^t$ is the $J\times K$ dimensional matrix with $X_{jk}$ as the $(j,k)$-th entry. We incorporate penalization to the functional effect $\beta(t)$ using the prior \eqref{eq:penaltyb} on the spline coefficient $\boldsymbol{b}$. To simplify the penalization prior, we apply the same reparametrization technique as in Section \ref{sec:2.2}. Keeping the same notation, the
Bayesian joint SoFR accounting for measurement error around the observed signal becomes:
\begin{equation}\label{eq:stansof_fpca}
    \begin{cases}
      \qquad \boldsymbol{Y} \sim l(\boldsymbol{Y},\boldsymbol{\delta};h_0,\boldsymbol{\eta})\;;\\
      \qquad \boldsymbol{\eta} = \eta_0\boldsymbol{J}_n+ \boldsymbol{\xi}\widetilde{\mathbf{X}}_r^t\boldsymbol{\widetilde{b}}_r+\boldsymbol{\xi}\widetilde{\mathbf{X}}_f^t\boldsymbol{\widetilde{b}}_f+\mathbf{Z}^t\boldsymbol{\gamma}\;;\\
      \qquad h_0(t)=\sum_{l=1}^L c_lM_l(t;\boldsymbol{k},\tau) \;;\\
      \qquad W_i(t)=\sum_{j=1}^J\xi_{ij}\phi_j(t) +\epsilon_i(t)\;;\\
      \\
      \qquad \xi_{ij} \sim N(\widehat{\xi}_{ij}, \lambda_j), \lambda_j \sim p(\lambda_j),  j=1,\ldots,J\;;\\
      \qquad \epsilon_i(t)\sim N(0,\sigma^2_\epsilon),\sigma^2_\epsilon \sim p(\sigma^2_\epsilon); \boldsymbol{c}\sim p(\boldsymbol{c})\;;\\
      \qquad \eta_0 \sim p(\eta_0);  \boldsymbol{\widetilde{b}}_r \sim p(\boldsymbol{\widetilde{b}}_r);\boldsymbol{\widetilde{b}}_f \sim p(\boldsymbol{\widetilde{b}}_f)\;;\\
      \qquad \boldsymbol{\gamma}\sim p(\boldsymbol{\gamma}); \sigma^2_b \sim p(\sigma^2_b)\;.\\
      
    \end{cases}       
\end{equation}
where $\widehat{\xi}_{ij}$ are the frequentist estimates of the scores using FPCA, $p(\lambda_j)$ and $p(\sigma^2_\epsilon)$ are non-informative priors for the variance components, such as $IG(0.001,0.001)$, though other priors could also be considered. The priors for $\boldsymbol{\widetilde{b}}_f$ and $\boldsymbol{\widetilde{b}}_r$ and $\mathbf{\gamma}$ are the same as in Section~\ref{sec:2}, while the prior on $\boldsymbol{c}$ is the Dirichlet prior with parameter $\boldsymbol{\alpha}=(1,\ldots,1)$ described in Section~\ref{sec:3}.


\subsection{Bayesian Joint Functional Modeling in Stan}

\subsubsection{Data preparation}
We first apply the frequentist FPCA to the observed functional covariate \texttt{data\$wmat} using the \texttt{refund::fpca.face} function in \texttt{R}:

\vspace{2mm}

\noindent\colorbox{backcolour}{
\begin{minipage}{\textwidth}
{ \ttfamily\footnotesize	 
\comcolor{// Fit the functional PCA model}

\vspace{-1.5mm}
fpca.fit = refund::\funcol{fpca.face}(Y=data\opcol{\$}wmat)
}
\end{minipage}
}\\

\vspace{-1.5mm}
The estimated eigenfunctions can be extracted from \texttt{fpca.fit\$efunctions}. Similar to Section \ref{sec:2}, we use the \texttt{mgcv::smooth.construct} function in \texttt{R} to construct the spline basis:

\vspace{2mm}

\noindent\colorbox{backcolour}{
\begin{minipage}{\textwidth}
{ \ttfamily\footnotesize	 
\comcolor{// Construct raw spline basis}

\vspace{-1.5mm}
splinecons = mgcv::\funcol{smooth.construct}(object, data, NULL)
}
\end{minipage}
}\\

\noindent where \texttt{object} is the \texttt{mgcv} term that specifies the spline type, parameter, and functional predictor. For example, for the cubic regression splines with $10$ degrees of freedom, as described in Section \ref{sec:2.3}, the \texttt{object} term would be \texttt{s(tmat, by=lmat*wmat, bs="cr", k=10)}. The spline basis $\boldsymbol{\psi}(t)$ can be extracted as:

\vspace{2mm}

\noindent\colorbox{backcolour}{
\begin{minipage}{\textwidth}
{ \ttfamily\footnotesize	 
\comcolor{// Extract the raw spline basis}

\vspace{-1.5mm}
Psi\_mat = splinecons\opcol{\$}X
}
\end{minipage}
}\\

The corresponding penalty matrix $\mathbf{S}$ can be extracted as:

\vspace{2mm}

\noindent\colorbox{backcolour}{
\begin{minipage}{\textwidth}
{ \ttfamily\footnotesize	 
\comcolor{// Extract the penalty matrix}

\vspace{-1.5mm}
S\_mat = splinecons\opcol{\$}S
}
\end{minipage}
}\\

We then calculate the $\mathbf{X}$ matrix in (\ref{eq:fpcapred}) by integrating the eigenfunction with the spline basis:

\vspace{2mm}

\noindent\colorbox{backcolour}{
\begin{minipage}{\textwidth}
{ \ttfamily\footnotesize	 
\comcolor{// Calculate the design matrix }\\
X\_mat\_t = \funcol{matrix}(nrow=J\_num, ncol=K\_num)   

for(j in 1:J\_num)\{

\qquad    for(k in 1:K\_num) \{
    
\qquad \qquad        X\_mat\_t[j,k] = \funcol{sum}(fpca.fit\opcol{\$}efunctions[,j] \opcol{*} Psi\_mat[,k]) \opcol{/} M\_num

\qquad \}

\vspace{-1.5mm}
\}
}
\end{minipage}
}\\

\noindent where \texttt{X\_mat\_t} refers to the transpose of $\mathbf{X}$ matrix, \texttt{J\_num, K\_num} are the corresponding values of $J, K$, and \texttt{M\_num} is the number of functional observations $M$.
To transform the matrix $\mathbf{X}$ to $\widetilde{\mathbf{X}}$, we first perform the spectral decomposition of the extracted penalty matrix \texttt{S\_mat}:

\vspace{2mm}

\noindent\colorbox{backcolour}{
\begin{minipage}{\textwidth}
{ \ttfamily\footnotesize	 
\comcolor{// Spectral decomposition of the design matrix}

\vspace{-1.5mm}
eigendecomp = \funcol{eigen}(S\_mat, symmetric = TRUE)
}
\end{minipage}
}\\

The transformation is done using the following code:

\vspace{2mm}

\noindent\colorbox{backcolour}{
\begin{minipage}{\textwidth}
{ \ttfamily\footnotesize	 
\comcolor{// Obtain the design matrix}\\
rank = splinecons\opcol{\$}rank

E = \funcol{rep}(1, \funcol{ncol}(X\_mat))

E[1:rank] = \funcol{sqrt}(eigendecomp\opcol{\$}value[1:rank]) 
\comcolor{// Square root of eigenvalues}

X\_mat\_t = X\_mat\_t \opcol{\%*\%} eigendecomp\opcol{\$}vectors

\vspace{-1.5mm}
X\_mat\_t = \funcol{t}(\funcol{t}(X\_mat\_t) \opcol{/} E) 
}
\end{minipage}
}\\

\noindent where \texttt{rank} is the rank of the penalty matrix, which is equal to the number of random effects basis. The transformed random effects design matrix $\mathbf{\widetilde{X}}_r$ can be extracted as:

\vspace{2mm}

\noindent\colorbox{backcolour}{
\begin{minipage}{\textwidth}
{ \ttfamily\footnotesize	 
\comcolor{// Extract the random effects design matrix}

\vspace{-1.5mm}
X\_mat\_r = \funcol{t}(X\_mat\_t[,1:rank])
}
\end{minipage}
}

\vspace{2mm}

\noindent and the transformed fixed effect design matrix $\mathbf{\widetilde{X}}_f$ is

\vspace{2mm}

\noindent\colorbox{backcolour}{
\begin{minipage}{\textwidth}
{ \ttfamily\footnotesize	 
\comcolor{// Extract the fix effect design matrix}

\vspace{-1.5mm}
X\_mat\_f = \funcol{t}(X\_mat\_t[,(rank \opcol{+} 1):\funcol{ncol}(X\_mat\_t)])
}
\end{minipage}
}\\


\subsubsection{Stan code}

We now describe the \texttt{Stan} code for our Bayesian joint SoFR model with time-to-event outcomes. We skip here the \texttt{data} and \texttt{parameter} blocks, though they are provided in the supplementary materials. Instead, we focus on the \texttt{model} block and identify the main changes from Sections~\ref{sec:2} and \ref{sec:3}. The \texttt{model} block is:

\vspace{2mm}


\noindent\colorbox{backcolour}{
\begin{minipage}{\textwidth}
{ \ttfamily\footnotesize	 
\comcolor{// Stan code model block}\\
\funcol{model} \{

\qquad    \comcolor{// Baseline and cumulative baseline hazard functions}
    
\qquad    \funcol{vector}[N\_num] bhaz = Mbasis \opcol{*} c;
    
\qquad    \funcol{vector}[N\_num] cbhaz = Ibasis \opcol{*} c;
    
\qquad    \comcolor{// Linear predictor}
    
\qquad    \funcol{vector}[N\_num] eta = \funcol{rep\_vector}(0.0, N\_num);

}
\end{minipage}
}\\

\noindent\colorbox{backcolour}{
\begin{minipage}{\textwidth}
{ \ttfamily\footnotesize

\qquad    eta += eta\_0 \opcol{+} xi \opcol{*} X\_mat\_f' \opcol{*} betaf \opcol{+} xi \opcol{*} X\_mat\_r' \opcol{*} betar \opcol{+} Z\_mat' \opcol{*} gamma;

\qquad    \comcolor{// Likelihood for the time-to-event outcome}
    
\qquad    for (n in 1:N\_num) \{
    
\qquad\qquad      if (cens[n] == 0) \{

\qquad\qquad\qquad target += \funcol{cox\_log\_lpdf}(Y[n] | eta[n], bhaz[n], cbhaz[n]);

\qquad\qquad \}       
        
\qquad\qquad       else if (cens[n] == 1) \{

\qquad\qquad\qquad target += \funcol{cox\_log\_lccdf}(Y[n] | eta[n], bhaz[n], cbhaz[n]);

\qquad\qquad\}    

\qquad\} 

\qquad    \comcolor{// Likelihood for the functional observation in FPCA}
    
\qquad    target += \opcol{-} N\_num \opcol{*} M\_num \opcol{*} log(sigma\_e) \opcol{-} sum((xi \opcol{*} Phi\_mat \opcol{-} M\_mat) \opcol{\textnormal{\^}}2) \opcol{/} (2 \opcol{*} sigma\_e \opcol{\textnormal{\^}}2);
    
\qquad    \comcolor{// Set prior for the FPCA scores xi}
    
\qquad    for(nj in 1:J\_num)\{
    
\qquad\qquad        target += \opcol{-} N\_num \opcol{*} \funcol{log}(lambda[nj]) \opcol{-} 

\qquad \qquad \qquad \funcol{sum}((xi[,nj] \opcol{-} xi\_hat[,nj]) \opcol{\textnormal{\^}}2) \opcol{/} (2 \opcol{*} lambda[nj] \opcol{\textnormal{\^}}2);
        
\qquad    \}

\qquad    \comcolor{// Other priors}

\qquad    \funcol{vector}[L\_num] alpha = \funcol{rep\_vector}(1, L\_num);

\qquad    target += \funcol{dirichlet\_lpdf}(c | alpha);
    
\qquad    target += \funcol{normal\_lpdf}(betar | 0, sigma);
    
\qquad    target += \funcol{inv\_gamma\_lpdf}(sigma\opcol{\textnormal{\^}}2|0.001,0.001);
    
\qquad    for(nj in 1:J\_num)\{
    
\qquad\qquad        target += \funcol{inv\_gamma\_lpdf}(lambda[nj]|0.001,0.001);
        
\qquad    \}
    
\qquad    target +=  \funcol{inv\_gamma\_lpdf}(sigma\_e\opcol{\textnormal{\^}}2|0.001,0.001);

\vspace{-1.5mm}
\}
}
\end{minipage}
}\\

The priors for other parameters \texttt{eta\_0}, \texttt{betaf}, and \texttt{gamma} are omitted in the \texttt{Stan} model, effectively imposing uniform priors as discussed in Section~\ref{sec:2}.

\subsection{Simulations}
\subsubsection{Data generating mechanism}

Simulation experiments follow the same scenarios described in Sections~\ref{sec:2.4} and \ref{sec:3.4}. The only difference is that the observed functional data, $\mathbf{W}$, are now simulated based on the estimated eigenfunctions from the data application but with added independent errors $\epsilon_i(t_j)\sim N(0,\sigma^2_\epsilon)$. We consider two cases, one with moderate noise, $\sigma_\epsilon = 5$, and one with large noise, $\sigma_\epsilon=10$. Details on simulations are provided in the supplementary materials.

\subsubsection{Competing methods}
 We compare the performance of our Bayesian \texttt{Stan} program with the results obtained using the \texttt{gam} function from the \texttt{R} package \texttt{mgcv}. For both Bayesian and frequentist methods, we use the cubic regression spline (\texttt{bs="cr"}) with $10$ degrees of freedom (\texttt{k=10}) to fit the model. As described in  Crainiceanu et al.\cite{crainiceanu2024functional}, one key distinction between the two approaches is that the frequentist approach conditions on $\xi_{ij}$, whereas the joint Bayesian approach incorporates the uncertainty associated with estimating these scores. 


\subsubsection{Results}
Table \ref{tab:fpca} provides the simulation results, which indicate  that the joint Bayesian algorithm: (1)  has similar performance in terms of relative integrated squared error (RISE); and (2) achieves superior coverage, especially when the noise is large, likely due to accounting for the variability of the score predictions. 

\begin{center}
\begin{table*}[]
\centering
\caption{\color{black}Simulation results comparing the two-step approach that conditions on FPCA scores with the joint Bayesian (\texttt{Stan}) methods for functional predictors with time-to-event outcomes for different number of observations, $n$,  and signal levels, $\tau$.  The median RISE, mean coverage rate of the 95\% credible/confidence intervals, and prediction accuracy are reported.}\label{tab:fpca}
\resizebox{0.9\textwidth}{!}{
\color{black}
\begin{tabular}{cccccccccccccc}
\hline
& & &
\multicolumn{3}{c}{\textbf{n=100}} & 
\multicolumn{3}{c}{\textbf{n=200}} &
\multicolumn{3}{c}{\textbf{n=300}} &
\multicolumn{2}{c}{\textbf{n=500}} \\
\cline{4-5} \cline{7-8} \cline{10-11} \cline{13-14} 
 & &  & Bayes & Freq &  & Bayes & Freq &  & Bayes & Freq &  & Bayes & Freq \\
  \hline  \\
   & tau = 1 & RISE & 4.739 & 4.665 &  & 2.742 & 2.725 &  & 1.8 & 1.922 &  & 1.297 & 1.356 \\ 
   &   & Coverage & (96.6) & (94.5) &  & (96.3) & (93) &  & (96.9) & (93.4) &  & (95.9) & (91.7) \\ 
   &   & Prediction & 3.103 & 3.478 &  & 1.89 & 2.057 &  & 1.295 & 1.44 &  & 0.911 & 1.024 \\ 
   & tau = 2 & RISE & 1.704 & 1.752 &  & 0.954 & 0.986 &  & 0.698 & 0.607 &  & 0.42 & 0.346 \\ 
Median   &   & Coverage & (94.5) & (90.8) &  & (95.7) & (92) &  & (96.3) & (93.8) &  & (97) & (95.2) \\ 
variance   &   & Prediction & 1.156 & 1.265 &  & 0.685 & 0.67 &  & 0.457 & 0.407 &  & 0.245 & 0.209 \\ 
   & tau = 3 & RISE & 0.913 & 0.958 &  & 0.508 & 0.383 &  & 0.315 & 0.255 &  & 0.221 & 0.179 \\ 
   &   & Coverage & (94.5) & (91.6) &  & (96.6) & (95.1) &  & (97.4) & (96.5) &  & (97.9) & (97) \\ 
   &   & Prediction & 0.703 & 0.661 &  & 0.331 & 0.27 &  & 0.198 & 0.173 &  & 0.115 & 0.105 \\ 
   & tau = 5 & RISE & 0.434 & 0.332 &  & 0.232 & 0.189 &  & 0.165 & 0.143 &  & 0.11 & 0.096 \\ 
   &   & Coverage & (95.5) & (94.3) &  & (97.8) & (97.3) &  & (97.9) & (97.1) &  & (98.1) & (98) \\ 
   &   & Prediction & 0.265 & 0.23 &  & 0.116 & 0.11 &  & 0.085 & 0.078 &  & 0.055 & 0.053 \\ 

    \hline
    \\
 & tau = 1 & RISE & 5.788 & 5.807 &  & 2.782 & 2.833 &  & 1.912 & 1.984 &  & 1.293 & 1.29 \\ 
 &   & Coverage & (97.6) & (94) &  & (97.2) & (92.3) &  & (97.6) & (91.5) &  & (97.7) & (90.7) \\ 
 &   & Prediction & 3.939 & 4.12 &  & 1.842 & 2.092 &  & 1.43 & 1.545 &  & 0.94 & 1.081 \\ 
 & tau = 2 & RISE & 1.771 & 1.853 &  & 0.999 & 1.042 &  & 0.815 & 0.777 &  & 0.503 & 0.37 \\ 
Large &   & Coverage & (97.2) & (91.7) &  & (96.7) & (89.6) &  & (96.5) & (89.2) &  & (97.7) & (93.2) \\ 
variance &   & Prediction & 1.261 & 1.348 &  & 0.786 & 0.859 &  & 0.664 & 0.602 &  & 0.361 & 0.277 \\ 
 & tau = 3 & RISE & 1.061 & 1.111 &  & 0.619 & 0.479 &  & 0.418 & 0.298 &  & 0.23 & 0.175 \\ 
 &   & Coverage & (96.2) & (88.3) &  & (96.1) & (91.3) &  & (97.3) & (94) &  & (98.5) & (96.8) \\ 
 &   & Prediction & 0.807 & 0.922 &  & 0.446 & 0.338 &  & 0.288 & 0.218 &  & 0.143 & 0.115 \\ 
 & tau = 5 & RISE & 0.578 & 0.449 &  & 0.259 & 0.192 &  & 0.186 & 0.141 &  & 0.126 & 0.094 \\ 
 &   & Coverage & (95.8) & (90.6) &  & (97.8) & (95.6) &  & (98.4) & (96.5) &  & (99.2) & (98.2) \\ 
 &   & Prediction & 0.464 & 0.327 &  & 0.171 & 0.142 &  & 0.129 & 0.107 &  & 0.08 & 0.065 \\ 
\hline

  \end{tabular}

  }
\end{table*}

\end{center}

 

\section{Bayesian Function-on-Scalar Regression}\label{sec:5}

In this section, we provide the step-by-step tutorial for implementing the Bayesian Function-on-Scalar Regression (FoSR) model using \texttt{Stan}. The procedure for fitting a generalized multilevel Bayesian FoSR was introduced in  Goldsmith et al.\cite{goldsmith2015generalized}; the primary purpose of this section is to provide a comprehensive software solution using \texttt{Stan}. 


\subsection{The Bayesian FoSR Model}\label{subsec:BayesFoSR}
Let $Y_i(t)$ be the functional response for subject $i=1,\ldots,n$ and at time points $t=t_1,\ldots,t_M$. FoSR assumes the following model for the functional response:
\begin{equation}\label{eq:FOSRmodel}
    Y_i(t) = \sum_{p=1}^P X_{ip}\beta_p(t)+e_i(t)\;,
\end{equation}
where $X_{ip}$ with $p=1,\ldots, P$ are the scalar predictors, and $\beta_p(t)$ are the corresponding domain-varying (or functional) coefficients. The first covariate is often the intercept, $X_{i1}=1$.  Two key differences between FoSR and  traditional regression are that: (1) the outcome, $Y_i(t)$, is multivariate and often high-dimensional; and (2) the residuals $e_i(t)$ are correlated across $t$ (points in the domain). To account for these data features, we assume that the residuals $e_i(t)$ have a zero-mean Gaussian Process (GP) distribution, but we also indicate how the software can be modified for non-Gaussian outcomes (e.g., binary, Poisson). Specifically, we assume that $e_i(t) = \sum_{j=1}^J \xi_{ij}\phi_j(t)+\epsilon_i(t)$,
where $\phi_1(t),\ldots,\phi_J(t)$ are the eigenfunctions and $\epsilon_i(t)$ are assumed to be independent $N(0,\sigma^2_\epsilon)$. With this notation, the model becomes:
\begin{equation}
    \begin{split}
        Y_i(t) &= \sum_{p=1}^P X_{ip}\beta_p(t)+ \sum_{j=1}^J \xi_{ij}\phi_j(t)  +\epsilon_i(t)\\
        &= \sum_{p=1}^P X_{ip} \sum_{k=1}^K b_{pk}\psi_k(t) + \sum_{j=1}^J \xi_{ij}\phi_j(t)  +\epsilon_i(t) \\
        &=  \sum_{k=1}^K (\sum_{p=1}^P X_{ip}b_{pk})\psi_k(t) + \sum_{j=1}^J \xi_{ij}\phi_j(t)  +\epsilon_i(t)\;.
    \end{split}
\end{equation}
Denote by $B_{ik} = \sum_{p=1}^P X_{ip}b_{pk}$, by $\boldsymbol{B}_i=(B_{i1},\ldots,B_{iK})^t$ the $K\times 1$ dimensional vector with $B_{ik}$ as the $k$-th entry,  by $\boldsymbol{\psi}(t)=\{\psi_1(t),...,\psi_K(t)\}$ the $1\times K$ dimensional vector of spline coefficients evaluated at time $t$, by $\boldsymbol{\Psi}$ the $M\times K$ dimensional matrix with the $m$ row equal to $\boldsymbol{\psi}(t_m)$, by $\boldsymbol{\phi}(t)=\{\phi_1(t),...,\phi_J(t)\}$ the $1\times J$ dimensional vector of spline coefficients evaluated at time $t$, and by $\boldsymbol{\Phi}$ the $M\times J$ dimensional matrix with the $m$ row equal to $\boldsymbol{\phi}(t_m)$. With this notation, we have that $\sum_{k=1}^K (\sum_{p=1}^P X_{ip}b_{pk})\psi_k(t)=\boldsymbol{\psi}(t)\boldsymbol{B}_i$, $\sum_{j=1}^J \xi_{ij}\phi_j(t)=\boldsymbol{\phi}(t)\boldsymbol{\xi}_i$, and $Y_i(t)=\boldsymbol{\psi}(t)\boldsymbol{B}_i+\boldsymbol{\phi}(t)\boldsymbol{\xi}_i+\epsilon_i(t)$. If $\boldsymbol{Y}_i = \{Y_{i}(t_1),\ldots,Y_i(t_M)\}^t$ is the $M\times 1$ dimensional vector of observed data for study participant $i$ and $\boldsymbol{\epsilon}_i=\{\epsilon_{i}(t_1),\ldots,\epsilon_i(t_M)\}^t$ is the $M\times 1$ vector of random errors, we have $\boldsymbol{Y}_i=\boldsymbol{\Psi}\boldsymbol{B}_i+\boldsymbol{\Phi}\boldsymbol{\xi}_i+\boldsymbol{\epsilon}_i$. Let $\boldsymbol{b}_p=(b_{p1},\ldots,b_{pK})^t$ denote the $K\times 1$ dimensional vector of spline coefficients for $\beta_p(t)$,  and $\mathbf{S}$ denote the smoothing penalty structure for $\boldsymbol{b}_p$ as described in Section~\ref{sec:2}. The Bayesian FoSR model can then be represented as
\begin{equation}\label{eq:stanfosr}
    \begin{cases}
       \qquad \boldsymbol{Y}_i = \boldsymbol{\Psi} \boldsymbol{B}_i+ \boldsymbol{\Phi}\boldsymbol{\xi}_i+\boldsymbol{\epsilon}_i\;;\\
       \qquad B_{ik} = \sum_{p=1}^P X_{ip}b_{pk}, k=1,\ldots,K\;;
      \\
       \qquad p(\boldsymbol{b_p}) \propto \exp(-\boldsymbol{b}_p^t\mathbf{S}\boldsymbol{b}_p/2\sigma_p^2), p=1,\ldots,P\;;\\
       \qquad \xi_{ij} \sim N(0, \lambda_j), \lambda_j\sim p(\lambda_j), j=1,\ldots,J\;;\\
       \qquad \epsilon_i(t_m)\sim N(0,\sigma_\epsilon^2), i=1,...,n, t=t_1,\ldots,t_M\;;\\
       \qquad \sigma_\epsilon^2 \sim p(\sigma^2_\epsilon), \sigma_p^2 \sim p(\sigma^2_p), p=1,\ldots,P\;.
    \end{cases}       
\end{equation}
We impose the normal smoothing priors $p(\boldsymbol{b}_p)$ by specifying them in the model. \color{black} One could apply the SVD techniques described in Sections~\ref{sec:2} and \ref{sec:3}, but 
we use the original spline basis without transformation because: (1) this approach was adopted in Goldsmith et al. \citep{goldsmith2015generalized}, and (2) using the untransformed basis simplifies interpretation and avoids the need to back-transform the estimated coefficients. \color{black}
The priors $p(\lambda_j)$, $p(\sigma^2_\epsilon)$, and $p(\sigma^2_p)$ are all non-informative priors on variance components; here we use inverse Gamma priors $IG(0.001,0.001)$, but other priors could also be considered. Notice that every function $\beta_p(t)$ has its own smoothing parameter, $\sigma^2_p$, which allows for fixed effects parameters to have different levels of smoothing (more or fewer degrees of freedom.) \color{black} In model \eqref{eq:stanfosr}, we assume that the functional responses are observed on a common set of grid points across all subjects. This assumption ensures that the $\boldsymbol{\Psi}$ matrix is identical across subjects, which simplifies matrix operations in Stan. However, the FoSR model is, in principle, applicable when functional responses are observed on subject-specific grids. Developing efficient Bayesian algorithms for such settings remains an important direction for future work. \color{black}

\subsection{Bayesian FoSR Implementation in Stan}\label{subsec:RFoSR}

\subsubsection{Data preparation}

As illustrated in Section \ref{sec:4}, the eigenfunctions $\boldsymbol{\Phi}$ can be obtained by applying FPCA to the functional response $Y_i(t)$ using the frequentist \texttt{refund::fpca.face} function. Assume that \texttt{data\$yindex} is a $M\times n$ index matrix with the $i$-th column being the sequence $(t_1,...,t_M)$. The spline basis $\boldsymbol{\Psi}$ and the corresponding penalty matrix $\mathbf{S}$ can be obtained using the \texttt{smooth.construct} function:

\vspace{2mm}

\noindent\colorbox{backcolour}{
\begin{minipage}{\textwidth}
{ \ttfamily\footnotesize	
\comcolor{// Construct the spline basis}\\
object = \funcol{s}(yindex, bs = \strcol{"cr"}, k = 10)

\vspace{-1.5mm}
splinecons = mgcv::\funcol{smooth.construct}(object, data, NULL)
}
\end{minipage}
}\\

\noindent where \texttt{object} is the mgcv term that specifies the spline information as described in Section \ref{sec:4}. Notice that the \texttt{object} here is different from other sections because we now have functional responses. The spline basis $\boldsymbol{\Psi}$ can be extracted as:

\vspace{2mm}

\noindent\colorbox{backcolour}{
\begin{minipage}{\textwidth}
{ \ttfamily\footnotesize	 
\comcolor{// Extract the spline basis}

\vspace{-1.5mm}
Psi\_mat = splinecons\opcol{\$}X
}
\end{minipage}
}\\

The penalty matrix $\mathbf{S}$ can be extracted as:

\vspace{2mm}

\noindent\colorbox{backcolour}{
\begin{minipage}{\textwidth}
{ \ttfamily\footnotesize	 
\comcolor{// Extract the penalty matrix}

\vspace{-1.5mm}
S\_mat = splinecons\opcol{\$}S
}
\end{minipage}
}\\

\subsubsection{Stan code}

We now describe the \texttt{Stan} code for the Bayesian FoSR. The \texttt{data} block is presented below:

\vspace{2mm}

\noindent\colorbox{backcolour}{
\begin{minipage}{\textwidth}
{ \ttfamily\footnotesize	 
\comcolor{// Stan code data block}\\
\funcol{data} \{

\qquad    \funcol{int}<lower=1> N\_num;   \comcolor{// Total number of participants}
    
\qquad    \funcol{int}<lower=1> M\_num;   \comcolor{// Total number of observed functional time point}
    
\qquad    \funcol{matrix}[N\_num, M\_num] Y\_mat;   \comcolor{// Functional response }
    
\qquad    \funcol{int}<lower=1> P\_num;   \comcolor{// Number of scalar predictors}
    
\qquad    \funcol{matrix}[N\_num, P\_num] X\_mat;   \comcolor{// Design matrix for the scalar predictor}
    
\qquad    \funcol{int}<lower=1> J\_num;   \comcolor{// Number of FPCA eigenfunctions}
    
\qquad    \funcol{matrix}[J\_num, M\_num] Phi\_mat;   \comcolor{// Matrix of FPCA eigenfunctions}
    
\qquad    \funcol{int}<lower=1> K\_num;   \comcolor{// Number of spline basis}
    
\qquad    \funcol{matrix}[K\_num, M\_num] Psi\_mat;   \comcolor{// Matrix of spline basis}
    
\qquad    \funcol{matrix}[K\_num, K\_num] S\_mat;    \comcolor{// Penalty matrix}

\vspace{-1.5mm}
\}
}
\end{minipage}
}\\

The \texttt{parameter} block is:

\vspace{2mm}

\noindent\colorbox{backcolour}{
\begin{minipage}{\textwidth}
{ \ttfamily\footnotesize	 
\comcolor{// Stan code parameter block}\\
\funcol{parameters} \{

\qquad    \funcol{matrix}[K\_num, P\_num] beta;   \comcolor{// Spline coefficients}
    
\qquad    \funcol{matrix}[N\_num, J\_num] xi;   \comcolor{// FPCA Scores}
    
\qquad    \funcol{real}<lower=0> sigma\_eps;   \comcolor{// Standard deviation of independent error}
    
\qquad    \funcol{vector}<lower=0>[P\_num] sigma;   \comcolor{// Smoothing parameter}
    
\qquad    \funcol{vector}<lower=0>[J\_num] lambda;   \comcolor{// FPCA eigenvalues}

\vspace{-1.5mm}
\}

}
\end{minipage}
}\\

The \texttt{model} block is:

\vspace{2mm}

\noindent\colorbox{backcolour}{
\begin{minipage}{\textwidth}
{ \ttfamily\footnotesize	 
\comcolor{// Stan code model block}\\
\funcol{model} \{

\qquad    \funcol{matrix}[N\_num, M\_num] mu;   

\qquad    \comcolor{// Fitted mean matrix}
    
\qquad    mu = X\_mat \opcol{*} beta\opcol{'} \opcol{*} Psi\_mat \opcol{+} xi \opcol{*} Phi\_mat;   

\qquad    \comcolor{// Log-likelihood for functional response}
    
\qquad    target += \opcol{-} N\_num \opcol{*} M\_num \opcol{*} \funcol{log}(sigma\_eps) \opcol{/} 2 \opcol{-} \funcol{sum}((mu \opcol{-} Y\_mat)  \opcol{\:\textnormal{\^}}2) \opcol{/} (2 \opcol{*} sigma\_eps \opcol{\:\textnormal{\^}}2);

\qquad    \comcolor{// Prior for the penalized spline coefficients}

\qquad    for(np in 1:P\_num)\{
    
\qquad\qquad      target +=  (\opcol{-} beta[,np] \opcol{*} S\_mat \opcol{*} beta[,np]\opcol{'}) \opcol{/} (2 \opcol{*} sigma[np]\opcol{\:\:\textnormal{\^}}2);
      
\qquad\qquad      target += \funcol{inv\_gamma\_lpdf}(sigma[np]\opcol{\:\textnormal{\^}}2|0.001,0.001);
      
\qquad    \}

\qquad    \comcolor{// Prior for the FPCA scores}
    
\qquad    for(nj in 1:J\_num)\{
    
\qquad\qquad      target += \opcol{-} N\_num \opcol{*} \funcol{log}(lambda[nj]) \opcol{/} 2 \opcol{-}  \funcol{sum}((xi[,nj]) \: \:\opcol{ \textnormal{\^}}2) \opcol{/} (2 \opcol{*} lambda[nj]\:\:\opcol{\textnormal{\^}}2);
      
\qquad\qquad      target += \funcol{inv\_gamma\_lpdf}(lambda[nj] \opcol{\textnormal{\^}}2|0.001,0.001);
      
\qquad    \}
    
\qquad    \comcolor{// Other priors}
    
\qquad    target +=  \funcol{inv\_gamma\_lpdf}(sigma\_eps \opcol{\textnormal{\^}}2|0.001,0.001);
    
\}

}
\end{minipage}
}
\vspace{2mm}

Since \texttt{xi} and the remaining part of \texttt{beta} have non-informative priors, we omit their explicit expressions in the \texttt{Stan} model.


\subsection{Simulations}\label{subsec:BFSoR_sim}

\subsubsection{Data generating mechanism}
 
Here we adapt the simulation setting from the previous sections and add one scalar predictor $X_i\sim N(20, 10)$ for $i=1,\ldots,n$ with $n=100, 300, 500, 700$. The functional responses are generated from the model
$Y_i(t) = X_i\beta(t) + W_i(t) +\epsilon_i(t)$,
where $\beta(t) = \{0.084 - (t-0.5)^2\}\times\tau$ and $\tau=0.5, 1, 2, 4$. The random effects $W_i(t)$ are generated as in Section \ref{sec:2.4} using the principal components, and $\epsilon_i(t)$ are generated independently from $N(0,5)$. As usual, the total number of time points $M$ is set to $50$ and we compare the relative integrated squared error (RISE) of the Bayesian approach with that of the frequentist results using the \texttt{mgcv} package.  

\subsubsection{Results}

The simulation results are presented in Table \ref{tab:fosr}. The Bayesian FoSR algorithm has a similar performance with \texttt{mgcv}. 

\begin{center}
\begin{table*}[]
\centering
\caption{Simulation results for FoSR. Results are structured as in all other simulation tables. \label{tab:fosr}}
\resizebox{0.9\textwidth}{!}{
\color{black}
 \centering
\begin{tabular}{ccccccccccccc}
\hline
& &
\multicolumn{3}{c}{\textbf{n=100}} & 
\multicolumn{3}{c}{\textbf{n=300}} &
\multicolumn{3}{c}{\textbf{n=500}} &
\multicolumn{2}{c}{\textbf{n=700}} \\
\cline{3-4} \cline{6-7} \cline{9-10} \cline{12-13} 
  &  & Bayes & Freq &  & Bayes & Freq &  & Bayes & Freq &  & Bayes & Freq \\ 
  \hline
 $\tau=0.5$ & RISE & 0.0085 & 0.0067 &  & 0.0032 & 0.0034 &  & 0.0022 & 0.0025 &  & 0.0018 & 0.0021 \\ 
   & Coverage & (99.85) & (95.11) &  & (99.61) & (94.44) &  & (99.2) & (93.88) &  & (98.93) & (93.49) \\ 
 $\tau=1$ & RISE & 0.0025 & 0.0029 &  & 0.0012 & 0.0017 &  & 0.0009 & 0.0015 &   & 0.0008 & 0.0013 \\ 
   & Coverage & (99.42) & (94.2) &  & (98.21) & (92.3) &  & (96.94) & (88.98) &  & (95.71) & (86.59) \\ 
 $\tau=2$ & RISE & 0.001 & 0.0015 &  & 0.0007 & 0.0012 &  & 0.0006 & 0.0011 &  & 0.0006 & 0.0011 \\ 
   & Coverage & (98.12) & (90.96) &  & (94.85) & (82.79) &  & (92.29) & (76.97) &  & (90.26) & (73.6) \\ 
 $\tau=4$ & RISE & 0.0006 & 0.0011 &  & 0.0006 & 0.001 &  & 0.0005 & 0.001 &  & 0.0005 & 0.0010 \\ 
   & Coverage & (91.49) & (79.42) &  & (84.56) & (67.37) &  & (78.75) & (60.88) &  & (73.75) & (55.5) \\ 
    \hline
    
  \end{tabular}

  }
\end{table*}

\end{center}

\section{NHANES Data Applications}\label{sec:6}
We apply the Bayesian functional regression programs to quantify the association between scalar and functional predictors and mortality in the National Health and Nutrition Examination Survey (NHANES). 
\color{black} The functional predictor is the minute-level average daily physical activity measured using accelerometers. 
The outcome  is a binary indicator of 5-year all-cause mortality.
Crainiceanu et al. \cite{crainiceanu2024functional} conducted a SoFR using the NHANES dataset to examine the association between physical activity and all-cause mortality.
\color{black} 
In this section, our Bayesian analysis results are compared to those reported in Crainiceanu et al.\cite{crainiceanu2024functional}.

\subsection{NHANES Physical Activity Study}\label{subsec_NHANES_PA}
 NHANES is a nationwide study conducted by the United States Centers for Disease Control and Prevention (CDC) to assess the health and nutritional status of adults and children in the United States \citep{mirel2013national}. It is conducted in two-year waves with approximately $10{,}000$ participants per wave. For the purpose of this paper, we focus on data collected from the NHANES 2011-2012 and NHANES 2013-2014 waves, which included up to seven days of continuous physical activity data collected using a tri-axial wrist-worn accelerometer \citep{leroux2024nhanes}. Accelerometry data was summarized by NHANES at the minute level using the ``Monitor Independent Movement Summary'' (MIMS) units \citep{johnintille2018}. The distribution of MIMS at each time point exhibits substantial skewness and a log-transformation was applied at the minute level, which resulted in more symmetric marginal distributions. Based on previous studies \citep{crainiceanu2024functional,cui2022fast,cui2021additive}, for every individual, physical activity data were averaged over available days of that individual at each minute, resulting in a $1{,}440$-dimensional vector of average log-MIMS values for every study participant. Processing was conducted using a similar pipeline to that described in the \texttt{rnhanesdata} package \citep{2018rnhanesdata}.  Mortality was determined by linking the NHANES data to death certificate records from the National Death Index, maintained by the National Center for Health Statistics (NCHS), through the end of 2019. 
 
In this case study, the outcome is a binary indicator of 5-year all-cause mortality. \color{black} Although the NHANES study provides additional covariates, we limited our adjustment to sociodemographic factors as in Crainiceanu et al\citep{crainiceanu2024functional}. \color{black} The predictors are the minute-level physical activity expressed in average log-MIMS, age, gender, race, body mass index (BMI), poverty-to-income ratio (PIR), coronary heart disease (CHD), and education level. The data after preprocessing can be downloaded from 
 \begin{center}
\url{http://www.ciprianstats.org/sites/default/files/nhanes/nhanes_fda_with_r.rds} 
\end{center}

The following code loads the dataset:

\vspace{2mm}

\noindent\colorbox{backcolour}{
\begin{minipage}{\textwidth}
{ \ttfamily\footnotesize	 
\comcolor{// Load the NHANES dataset}

\vspace{-1.5mm}
nhanes\_use = \funcol{readRDS}(\strcol{"nhanes\_fda\_with\_r.rds"})
}
\end{minipage}
}\\

The NHANES 2011–2012 and 2013–2014 waves include a total of $12{,}610$ participants with accelerometry data. Participants with missing outcome  ($3{,}897$ participants) and scalar covariates (an additional $1{,}107$ participants) were excluded. One additional participant was excluded because their coronary heart disease (CHD) status was recorded as ``Refused''. After these exclusions, the final dataset used for this analysis included $7{,}605$ participants. The following code provides the data exclusion process from the \texttt{nhanes\_use} data set:

\vspace{2mm}

\noindent\colorbox{backcolour}{
\begin{minipage}{\textwidth}
{ \ttfamily\footnotesize	 
\comcolor{\# Names of the scalar predictors}\\
pred.names = c(\strcol{"age","gender","race","BMI","PIR","CHD"}\\

\vspace{-3mm}
\qquad\qquad\qquad\qquad \strcol{"education"})

\comcolor{\# Only include participants with fully observed scalar predictors and outcome}\\
for(cova in pred.names)\{

    \qquad nhanes\_use = nhanes\_use[\opcol{!}is.na(nhanes\_use[cova]),]
  
\}

nhanes\_use = nhanes\_use[\opcol{!}\funcol{is.na}(nhanes\_use\opcol{\$}event),]

nhanes\_use = nhanes\_use[-\funcol{which}(nhanes\_use\opcol{\$}CHD==\strcol{"Refused"}),]

\comcolor{\# Construct the grid matrix and time point matrix for functional predictor}\\
nhanes\_use\$lmat = \funcol{I}(\funcol{matrix}(1/1440, ncol=1440, nrow=\funcol{nrow}(nhanes\_use))) 

nhanes\_use\$tmat = \funcol{I}(\funcol{matrix}(1:1440, ncol=1440, nrow=\funcol{nrow}(nhanes\_use), byrow=\funcol{TRUE}))

}
\end{minipage}
}\\

\subsection{Frequentist Approach Based on Mixed Effects Representation}\label{sec:6.2}
 We fit the scalar-on-function regression model described in \eqref{eq:linearpred} for a binary outcome $Y_i$ (an indicator of death during the follow-up period), functional predictor $W_i(t)$ (average physical activity at time $t$ from midnight), and covariates $\mathbf{Z}_i$ (described in Section~\ref{subsec_NHANES_PA}). We first fit SoFR using the frequentist approach \cite{goldsmith2011penalized,crainiceanu2024functional} and implemented using the \texttt{mgcv::gam} function; as described in Crainiceanu et al.\cite{crainiceanu2024functional}, this can be implemented in \texttt{refund::pfr} \citep{goldsmith_refund_2024}. Since the functional covariate is periodic (midnight is both $0$ and $24$), we model the functional effect, $\beta(t)$, as periodic using a cyclic cubic penalized spline. The frequentist model based on mixed effects representation of SoFR can be fit using the following code:

\vspace{2mm}

\noindent\colorbox{backcolour}{
\begin{minipage}{\textwidth}
{ \ttfamily\footnotesize	 
\comcolor{// Frequentist model based on mixed effects representation}\\
\funcol{library}(mgcv)

fit\_freq\_b = \funcol{gam}(event \opcol{$\sim$} age \opcol{+} gender \opcol{+} race \opcol{+} BMI \opcol{+} PIR \opcol{+} CHD \opcol{+}

\hspace{0.3in}  education \opcol{+} \funcol{s}(tmat,  by=lmat\opcol{*}MIMS, bs=\strcol{"cc"}, k=10),

\vspace{-1.5mm}
\hspace{0.3in}  data=nhanes\_use,  family=\funcol{binomial}())
}
\end{minipage}
}\\

The estimated functional and scalar coefficients as well as their confidence intervals can be obtained through the \texttt{mgcv::plot.gam} and \texttt{mgcv::summary} functions:

\vspace{2mm}

\noindent\colorbox{backcolour}{
\begin{minipage}{\textwidth}
{ \ttfamily\footnotesize	 
\comcolor{// Estimated functional coefficient}

\funcol{plot.gam}(fit\_freq\_b)

\comcolor{// Estimated scalar coefficients}

\vspace{-1.5mm}
\funcol{summary}(fit\_freq\_b)
}
\end{minipage}
}\\

\begin{figure*}
    \centering
    \caption{Estimated functional effect for the scalar-on-function regression for Bayesian (left) and Frequentist (right) methods.  Darker gray shaded area bordered by dashed lines: pointwise $95$\% confidence/credible interval. Lighter gray shaded area bordered by dotted lines: CMA $95$\% confidence/credible interval.}
     \includegraphics[width=0.9\textwidth]{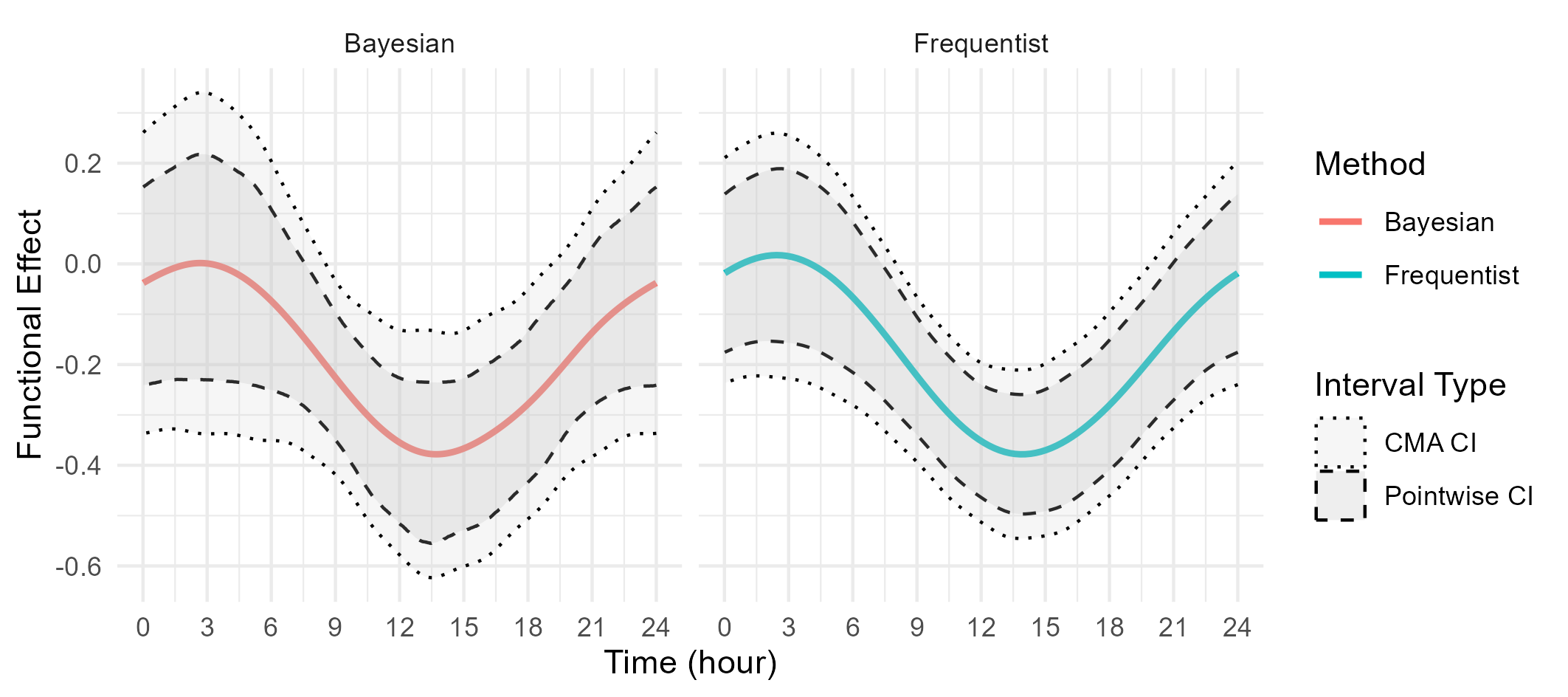}
    \label{Fig:1}
\end{figure*}

Results for the functional coefficient are presented in Figure \ref{Fig:1} (right panel, labeled ``Frequentist''). In Table \ref{table:1}, we present the estimated functional regression coefficients.
\begin{table*}[]
    \centering
    \scriptsize
    \renewcommand{\arraystretch}{1.1}  
    \caption{Estimated coefficients for scalar predictors in the NHANES example with corresponding 95\% confidence/credible intervals for the frequentist and Bayesian implementation.}
    \label{table:1}
    \resizebox{0.7\textwidth}{!}{%
        \begin{tabular}{|l|c|c|}
            \toprule
            \textbf{Scalar Covariate} & \textbf{Bayesian Estimate} & \textbf{Frequentist Estimate}  \\ 
            \midrule
            Age  & 0.063 ( 0.054 , 0.073 )  & 0.063 ( 0.054 , 0.072 )\\ 
                
            Gender (Female)  & -0.068 ( -0.280 , 0.148 ) & -0.066 ( -0.278 , 0.146 ) \\ 
                  
            BMI   & -0.023 ( -0.039 , -0.007 ) & -0.023 ( -0.039 , -0.007 )  \\ 
                
            PIR   & -0.165 ( -0.243 , -0.088 ) & -0.163 ( -0.240 , -0.087 ) \\        
            \midrule
            Race &&\\
            
            \qquad Other Hispanic   & 0.115 ( -0.472 , 0.720 ) & 0.105 ( -0.480 , 0.690 ) \\ 
                                    
            \qquad Non-Hispanic White  & 0.649 ( 0.201 , 1.127 ) & 0.626 ( 0.164 , 1.089 )  \\ 
                                         
            \qquad Non-Hispanic Black  & 0.454 ( -0.023 , 0.955 ) & 0.432 ( -0.050 , 0.914 ) \\ 
                                        
            \qquad Non-Hispanic Asian   & -0.259 ( -0.921 , 0.400 ) & -0.256 ( -0.918 , 0.405 ) \\ 
                                       
            \qquad Other Race  & 0.734 ( -0.061 , 1.500 ) & 0.738 ( -0.037 , 1.514 ) \\ 
                 
            \midrule
            CHD &&\\
            
            \qquad Yes  & 0.568 ( 0.255 , 0.872 ) & 0.569 ( 0.263 , 0.874 ) \\ 
                       
            \qquad Don't know  & 1.331 ( 0.335 , 2.301 ) &  1.324 ( 0.376 , 2.271 ) \\ 
                                
            \midrule
            Education &&\\
            
            \qquad High school equivalent  & -0.146 ( -0.429 , 0.138 ) &  -0.143 ( -0.424 , 0.137 )  \\ 
                                                
            \qquad More than high school  & -0.370 ( -0.642 , -0.094 ) & -0.368 ( -0.639 , -0.097 ) \\ 
                                                 
            \qquad Don't know  & 2.199 ( -0.428 , 5.592 ) & 1.811 ( -0.692 , 4.315 ) \\ 
                                      
            \bottomrule
        \end{tabular}
    }
\end{table*}

\subsection{Bayesian Approach Using Stan}\label{sec:6.3}

\subsubsection{Bayesian model fit}

We now analyze the same model using the Bayesian \texttt{Stan} program introduced in this paper on the same dataset. Our \texttt{R} package \texttt{refundBayes} was developed to have a similar syntax to that in \texttt{mgcv::gam}, though it uses Bayesian posterior inference and a few additional arguments that are specific to the \texttt{Stan} implementation. The eventual goal is for the user to simply run the program without implementing it from scratch. 

The \texttt{bfrs} function uses the same data format as the \texttt{mgcv::gam} function, including the spline construction.  The following code shows how \texttt{refundBayes} package fits the Bayesian SoFR using \texttt{Stan}:

\vspace{2mm}

\noindent\colorbox{backcolour}{
\begin{minipage}{\textwidth}
{ \ttfamily\footnotesize	 
\comcolor{// Bayesian approach using refundBayes package}\\
\funcol{library}(refundBayes)

fit\_bfrs = \funcol{bfrs}(event \opcol{$\sim$} age \opcol{+} gender \opcol{+} race \opcol{+} BMI \opcol{+} PIR \opcol{+} CHD \opcol{+} 

\hspace{0.3in} education \opcol{+}  \funcol{s}(tmat,  by=lmat\opcol{*}MIMS, bs=\strcol{"cc"}, k=10),  

\hspace{0.3in} data=nhanes\_use,  family=\funcol{binomial}(), joint\_FPCA=F, 

\vspace{-1.5mm}
\hspace{0.3in} n.iter=15000,  n.warmup=5000, n.knots=3)
}
\end{minipage}
}\\

For the \texttt{bfrs} function, the \texttt{data} and the \texttt{family} arguments specify the dataset and the outcome type, which have the same meaning as in the \texttt{gam} function. The \texttt{joint\_FPCA} argument specifies whether to simultaneously perform FPCA within a joint model, as described in Section \ref{sec:5}. The \texttt{n.iter}, \texttt{n.warmup}, and \texttt{n.knots} arguments are the total number of iterations, the burn-in value, and the number of chains (and knots for parallel computing) for posterior samples, respectively. The \texttt{Stan} program in this example used $15{,}000$ posterior samples with $5000$ burn-in iterations using $3$ different chains.

\color{black}We use $k=10$ number of basis in this example for illustrative purposes. In practical Bayesian analysis, the choice of the number of basis functions involves a trade-off between model flexibility and computational cost. Too few basis functions may fail to capture the true functional signal, while too many can result in unstable estimates and higher computational burden, even when penalization on the spline coefficients is applied. We recommend selecting the basis dimension by considering the complexity of the underlying functions, the resolution of the observed data, and the sample size. As a general guideline, 30–40 basis functions are often sufficient for moderately smooth functional data observed on dense grids \citep{xiao2016fast}. \color{black}

\subsubsection{Results}

The function \texttt{plot.bfrs(fit\_bfrs)} displays the fitted functional coefficients with both the pointwise and the correlation and multiplicity adjusted (CMA) credible intervals. The CMA credible interval is constructed in a similar way as the CMA confidence interval described in Section 2.4.3 of Crainiceanu et al \cite{crainiceanu2024functional}. Specifically, denote by $\beta^q(t)$ the $q$-th posterior sample of the functional effect, $\beta(t)$. Then, we can calculate the posterior samples from the max statistic
$d^q = \max\{|\beta^q(t) - \widehat{\beta}(t)|/\widehat{S}(t):t=t_1,\ldots,t_M \}$
where $\widehat{\beta}(t_i)$  and  $\widehat{S}(t_i)$ is the mean and standard deviation of the posterior sample $\{\beta^q(t_i)\}_{q=1}^Q$, respectively. Then, the upper and lower bound of the CMA credible interval can be calculated as $\widehat{\beta}(t) \pm q_{d,1-\alpha} \times \widehat{S}(t)$,
where $q_{d,1-\alpha}$ is the $100(1-\alpha)$ percentile of the distribution of $\{d^q, q=1,...,Q\}$.

The Bayesian estimated functional coefficient and $95$\% credible intervals are displayed in the left panel of Figure \ref{Fig:1}. The pointwise credible intervals are displayed as a darker shade of gray bordered by dashed lines, while the CMA credible intervals are displayed as a lighter shade of gray bordered by dotted lines. Compared with the estimators for the frequentist approach displayed in the right panel of the same figure, we conclude that the shape and length of confidence intervals are largely comparable, though the Bayesian credible intervals tend to be about $21$\% wider. 
This could be due to accounting for the uncertainty of the smoothing parameter, though more investigation would be necessary to identify the sources of this variation. For both Bayesian and frequentist approaches, the CMA credible interval is roughly $32$\% wider than their pointwise counterparts. 
Results for the scalar coefficients are presented in the Table \ref{table:1} (labeled “Bayes”), which indicates an excellent agreement with the frequentist results. 

\subsubsection{Bayesian model diagnosis}
\color{black}
Assessing the convergence of Markov Chain Monte Carlo (MCMC) algorithms is a critical step in Bayesian modeling. One commonly used diagnostic tool for evaluating convergence is the traceplot, which displays the sampled values of a parameter across iterations of the MCMC algorithm. An ideal traceplot will show convergence with the parameter values oscillating around the mode of the posterior distribution.\citep{gunn2024practical} Visually inspection using the \texttt{traceplot} function from \texttt{R} \texttt{rstan} package is displayed in Figure 1 of the Supplementary materials and indicates good convergence. 

\color{black}



\section{Discussion}\label{sec:7}
We introduced a Bayesian \texttt{Stan} program for scalar-on-function, functional Cox, and function-on-scalar regression models. This work builds upon the core ideas for functional regression described in Crainiceanu et al.\cite{crainiceanu2024functional}: model functional effects parametrically or nonparametrically using splines, penalize the spline coefficients, and conduct inference in the resulting mixed effects models. These ideas have a long history with previous Bayesian implementations \citep{crainiceanuwinbugs,crainiceanu2010bayesian} in \texttt{WinBUGS} \citep{lunn2000winbugs}. However, the emergence of the powerful \texttt{Stan} \citep{carpenter2017stan}  Bayesian package provides substantial computational improvements and seamless integration with \texttt{R} \citep{R}. Simulation results indicate that our specific implementation has reasonable inferential properties and compares well with existing approaches.

More work will need to be done to address computational stability and sensitivity to prior choices, but the approach is fully reproducible; see the supplementary material. \color{black} When repeated measures of functional data are available (e.g., over seven days in our example), investigating the variability across those measurements is of interest. One natural approach is to apply multilevel functional principal component analysis (MFPCA). Our Bayesian joint modeling framework in Section 4 can be easily extended to MFPCA settings, although a full investigation of its statistical properties is left for future work. \color{black}





\allowdisplaybreaks


%
%




\def\spacingset#1{\renewcommand{\baselinestretch}%
{#1}\small\normalsize} \spacingset{0.5}

\providecommand{\bmsection}{\section}
\bibliographystyle{Chicago}
\bibliography{Bibliography}

\end{document}